\documentclass{article}

\usepackage{PRIMEarxiv}

\usepackage[utf8]{inputenc} 
\usepackage[T1]{fontenc}    
\usepackage{hyperref}       
\usepackage{url}            
\usepackage{booktabs}       
\usepackage{amsfonts}       
\usepackage{nicefrac}       
\usepackage{lipsum}
\usepackage{fancyhdr}       
\usepackage{graphicx}       

\usepackage{tikz}
\usepackage{nicematrix}
\usepackage{mdframed}
\usepackage{subcaption}
\usepackage{paralist}
\usepackage{multirow}
\usepackage{amsmath}
\usepackage{float}
\usepackage{inconsolata}
\usepackage{latexsym}
\usepackage[numbers]{natbib}
\graphicspath{{media/}}     

\title{Evaluating Perspectival Biases in Cross-Modal Retrieval
\thanks{Published in \emph{Findings of the Association for Computational Linguistics: ACL 2026}, pages 36018--36049. \url{https://aclanthology.org/2026.findings-acl.1795/}}
}

\author{
    \textbf{Teerapol Saengsukhiran}\textsuperscript{1}\thanks{These authors contributed equally as co-first authors.} , 
    \textbf{Peerawat Chomphooyod}\textsuperscript{1}\footnotemark[1] , 
    \textbf{Narabodee Rodjananant}\textsuperscript{1}\footnotemark[1] , \\
    \textbf{Chompakorn Chaksangchaichot}\textsuperscript{1,2}, 
    \textbf{Patawee Prakrankamanant}\textsuperscript{1},
    \textbf{Witthawin Sripheanpol}\textsuperscript{1}, \\ \textbf{Pak Lovichit}\textsuperscript{1}, \textbf{Sarana Nutanong}\textsuperscript{3}, and \textbf{Ekapol Chuangsuwanich}\textsuperscript{1}\thanks{Corresponding Author (Email: ekapolc@cp.eng.chula.ac.th)}
    \vspace{0.5cm}
    \\
    \textsuperscript{1}Department of Computer Engineering, Chulalongkorn University, Bangkok, Thailand 10330\\
    \textsuperscript{2}VISAI.AI, Bangkok, Thailand 10110\\
    \textsuperscript{3}School of Information Science and Technology, VISTEC, Rayong 21210
}

\begin{document}
\maketitle

\begin{abstract}
Multimodal retrieval systems are expected to operate in a semantic space, agnostic to the language or cultural origin of the query.
In practice, however, retrieval outcomes systematically reflect perspectival biases: deviations shaped by linguistic \textbf{prevalence} and \textbf{cultural} associations.
We introduce \textbf{the Cross-Cultural, Cross-Modal, Cross-lingual Multimodal (3XCM)} benchmark to isolate these effects.
Results from our studies indicate that, for image-to-text retrieval, models tend to favor entries from prevalent languages over those that are semantically faithful.
For text-to-image retrieval, we observe a consistent \emph{``tugging effect''} in the joint embedding space between semantic alignment and language-conditioned cultural association.
When semantic representations are insufficiently resolved, particularly in low-resource languages, similarity is increasingly governed by culturally familiar visual patterns, leading to systematic association bias in retrieval.
Our findings suggest that achieving equitable multimodal retrieval necessitates targeted strategies that explicitly decouple language from culture, rather than relying solely on broader data exposure. 
This work highlights the need to treat linguistic and cultural biases as distinct, measurable challenges in multimodal representation learning.
The 3XCM benchmark is publicly available at \url{https://huggingface.co/datasets/Chula-AI/association_bias_benchmark}.
\end{abstract}

\section{Introduction}
\begin{figure}[!ht]
\centering
\includegraphics[width=\columnwidth]{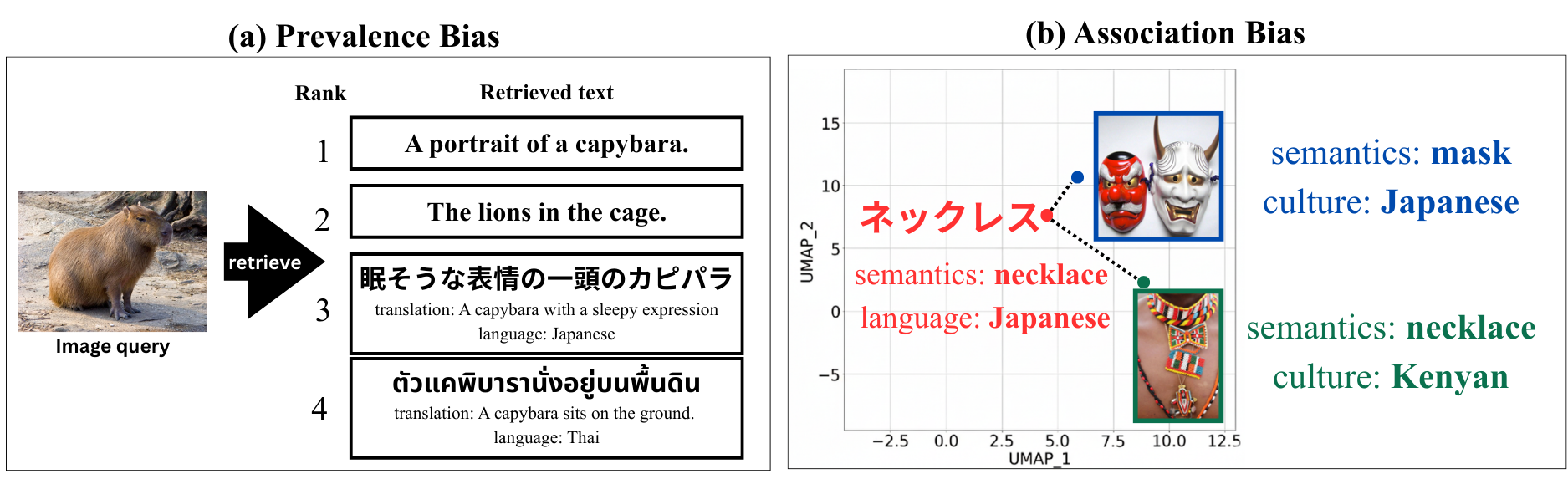}
\caption{Two Forms of Perspectival Biases. (a) \textbf{Prevalence bias}: an image query favors high-resource languages. A model places English results above semantically equivalent Japanese and Thai captions. (b) \textbf{Association Bias}: Given a Japanese text query for ``necklace'', a model places a culturally proximate image (Japanese masks) closer to the query than a semantically correct one (Kenyan necklace).}
\label{fig:bias_definition}
\vspace{-1.5em}
\end{figure}

As \citet{nietzsche1887genealogy} observed, \emph{there is only a perspective seeing, only a perspective knowing''}; put differently, there is \emph{no view from nowhere}. %
Large models inherit this perspectival character from training data, where representations depend on frequency and co-occurrence. %
Consequently, their latent spaces deviate from the expected robust, language-agnostic semantics, skewing retrieval toward linguistic prevalence or cultural associations over true relevance. %
Figure~\ref{fig:bias_definition} illustrates this in image-to-text and text-to-image retrieval, highlighting the need to quantify such effects for consistent cross-lingual, cross-cultural performance. %

Multimodal retrieval aligns text and images via paired supervision in early models like CLIP \citep{DBLP:journals/corr/abs-2103-00020} or implicit pretraining in recent Multimodal Large Language Models (MLLMs) \citep{bai2025qwen25vltechnicalreport,comanici2025gemini}. %
Yet, language and cultural biases remain underexplored, especially given English-centric datasets like LAION \citep{schuhmann2022laion5bopenlargescaledataset} and WebLi \citep{chen2023pali}. %
While culturally specific images may carry native-language alt-text, e.g., Catalan descriptions for ``coca de recapte'' \citep{olondriz2021foodi}. This fosters emergent multilingualism but risks spurious correlations, favoring ``expected'' languages. %

A major barrier is the lack of targeted benchmarks and metrics. %
To address this, we develop an evaluation framework isolating two perspectival biases across retrieval directions. %
\textbf{For image-to-text} (lacking linguistic cues), we assess \emph{prevalence bias} via the proposed Discounted Language Bias Kullback--Leibler Divergence (DLBKL), extending LBKL \citep{laosaengpha2025mitigatinglanguagebiascrosslingual} to penalize high-resource languages dominating top ranks (Figure~\ref{fig:bias_definition}a). 
\textbf{For text-to-image} (with linguistic/cultural cues), we measure \emph{association bias} using a parallel, cross-cultural dataset to disentangle semantic fidelity from cultural proximity (Figure~\ref{fig:bias_definition}b). %
Extending this, we test the ability of explicit cultural descriptors (CDs) to override association bias, examining the ``tugging effect'' where models struggle to prioritize CDs (e.g., ``Japanese train'' in Thai) over implicit query-language associations. %

Using these tools, we compare biases in MLLM-adapted retrievers against those with explicit cross-lingual alignment \citep{chen-etal-2023-mclip, carlsson-etal-2022-cross}. %
Findings expose biases in both directions, including CDs' limited efficacy: models follow instructions but fall back to language-linked visuals when cultural semantics are unresolved. %

Our contributions:
\begin{inparaenum}[(i)]
\item DLBKL as a supplementary, rank-sensitive extension to LBKL for measuring prevalence bias in multilingual pools; %
\item the 3XCM benchmark, parallel across cultures/languages for association bias; %
\item analysis of CDs' impact on retrieval, revealing persistent biases despite explicit guidance. %
\end{inparaenum}

\section{Related Works}

\subsection{Language Bias in Multimodal Retrievers}
\label{sec:mr_bias}
Language bias in multimodal retrieval refers to performance disparities where high-resource languages dominate rankings for semantically equivalent queries. 
\citet{osmulski2025miracl} report that modern retrievers exhibit significant variation in NDCG scores \cite{ndcg} across languages, indicating that their effectiveness depends on resources and script types. 

To quantify these disparities, fairness-aware metrics like exposure parity \cite{ndcg} have been adapted, yet \citet{adewumi2024fairnessbiasmultimodalai} emphasizes the lack of dedicated language-focused protocols. 
\citet{laosaengpha2025mitigatinglanguagebiascrosslingual} introduced LBKL, a distributional measure of divergence for text modality bias. 
While technically extensible to multimodal retrieval, such metrics ignore retrieval rankings in the bias measurement.

\subsection{Cultural Benchmarks} 

While recent benchmarks \cite{liu-etal-2021-visually,romero2024cvqa,nayak-etal-2024-benchmarking,li2025ravenea} have made progress in evaluating cultural reasoning and knowledge, they primarily focus on culture-specific concepts or broad facets.
In contrast, the assessment of bias discussed in Section~\ref{sec:mr_bias} requires a basic-level, parallel corpus of universal concepts to strictly disentangle semantic relevance from cultural association in retrieval rankings.
We provide a detailed comparison to other datasets in Appendix~\ref{apd:compare_dataset}.

\section{Methodology}

This section outlines the framework developed to investigate perspectival bias in multilingual, multimodal retrieval systems. 
We first state our guiding research questions and subsequently detail the experiments addressing them in Sections~\ref{sec:rq1}, \ref{sec:rq2}, and~\ref{sec:rq3}.
Our investigation is organized into three specific research questions as follows:

\begin{compactenum}[\bf RQ1]
    \item \textbf{[Image$\rightarrow$Text]:} Effect of \emph{prevalence bias}. To what extent do models favor high-resource languages over semantically equivalent captions in other languages?
    
    \item \textbf{[Text$\rightarrow$Image]:} Effect of \emph{culture-language association bias}. To what extent do models prioritize culture associated with the query language over semantically faithful results?
    
    \item \textbf{[Text$\rightarrow$Image]:} Effect of \emph{culture-language association bias under explicit cultural description}. To what extent can models adhere to explicit cultural descriptors (e.g., country mentions) against the \emph{``tugging effect''} of inherent language-cultural association biases?
\end{compactenum}

\subsection{Image-to-Text Retrieval (RQ1)}
\label{sec:rq1}

To assess prevalence bias, we measure the divergence between expected and observed language distributions. We adopt the Language Bias Kullback–Leibler (LBKL) divergence \cite{laosaengpha2025mitigatinglanguagebiascrosslingual}:
\begin{equation}
\small
\text{LBKL} = \frac{\sum\limits_{i=1}^{q} \left[ P_{\text{A}}(x) \log \frac{P_{\text{A}}(x)}{Q_{\text{A}}(x)} + P_{\text{B}}(x) \log \frac{P_{\text{B}}(x)}{Q_{\text{B}}(x)} \right]}{q}
\label{lbkl equation}
\end{equation}
Since LBKL is rank-agnostic, we propose \textbf{Discounted LBKL (DLBKL)} to prioritize penalizing bias in higher ranks more than lower ranks. We apply a logarithmic discount $w(i) = 1/\log_2(i+1)$ to the observed language proportion $Q'_l(x)$:
\begin{equation}
Q'_l(x) = \frac{\sum_{i=1}^k w(i) \cdot \mathbb{I}(\text{doc}_i \text{ is } l)}{\sum_{i=1}^k w(i)}
\label{dlbkl equation}
\end{equation}

\begin{figure}[htbp]
  \centering
  \includegraphics[width=0.7\columnwidth]{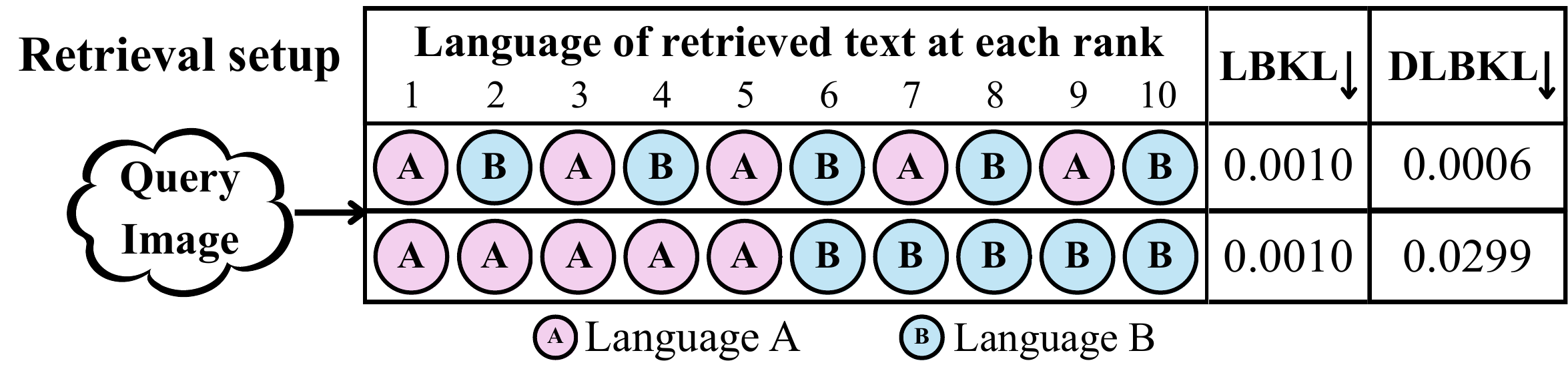}
  \caption{Illustration of how DLBKL, unlike the rank-agnostic LBKL, assigns a higher bias score to lists in which high-resource languages dominate the top ranks, under the setup that the prior P is a uniform distribution}
  \label{fig:lbkl_vs_dlbkl}
  \vspace{-2mm}
\end{figure}

DLBKL is computed by substituting $Q'_l(x)$ for $Q_l(x)$ in equation~\ref{lbkl equation}. 
This formulation can be generalized to support cases where the bias assessment involves more than two languages by measuring divergence over the full distribution. 
\emph{It penalizes models where high-resource languages disproportionately occupy top ranks} relative to the prior $P$. 
Crucially, in parallel datasets where semantic relevance is uniform across languages, this formulation isolates prevalence bias without compromising retrieval accuracy (see Figure~\ref{fig:lbkl_vs_dlbkl} and Appendix~\ref{apd:dlbkl_details}).

\subsection{Text-to-Image Retrieval (RQ2 \& RQ3)}
\label{sec:text_to_image}

To quantify the degree to which models prioritize cultural association over semantic fidelity, a phenomenon we illustrate in Figure \ref{fig:bias_definition}(b), a benchmark with a parallel structure in its cultural dimension is necessary. 
To the best of our knowledge, no such benchmark exists, so we make two primary contributions. 
First, we construct and introduce the Cross-Cultural, Cross-Modal, Cross-lingual Multimodal (3XCM) benchmark, a novel dataset designed specifically for this purpose. 
Second, we propose the Self-Preference Cultural Bias Score (SP), a new metric for explicitly measuring this form of bias.

\subsubsection{The 3XCM Dataset Benchmark}

\begin{figure}[h]
  \centering
\includegraphics[width=1\textwidth]{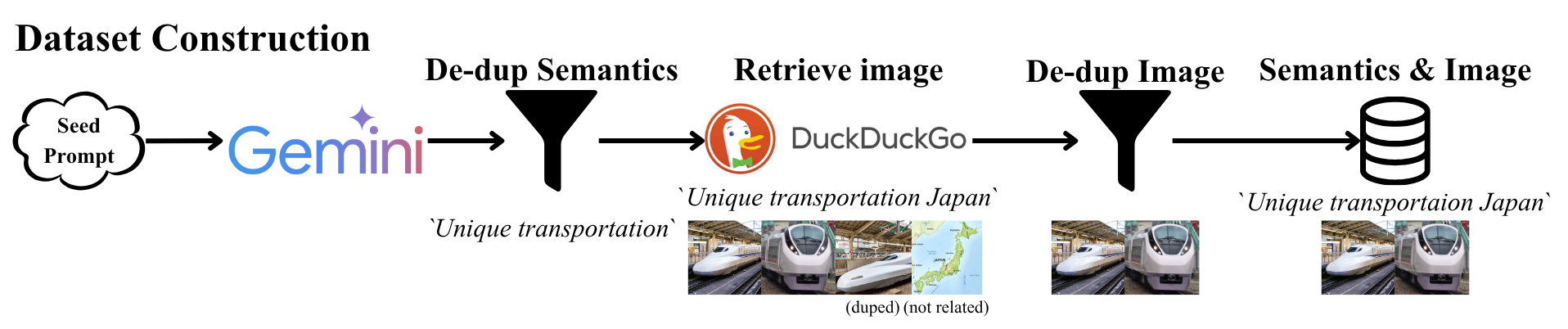}
  \caption{Overview of the XCM dataset creation process, designed to produce a benchmark with parallelism across semantics, cultures, and languages.}
  \label{fig:dataset}
\end{figure}

To evaluate association bias, we constructed the 3XCM benchmark, a challenge set designed to elicit culture-language association bias. This evaluation requires strict cultural parallelism at the concept level (e.g., "school" or "wedding" across all cultures), a structural feature that existing datasets \cite{liu-etal-2021-visually, romero2024cvqa, nayak-etal-2024-benchmarking} do not provide.\footnote{Research release only (CC BY-NC-SA 4.0). Ethical review required for production use. \label{fn:dataset_cc}}
The process involved two primary stages: (i) gathering a corpus of culturally diverse images and (ii) structuring these images into a triplet-based evaluation set.

The image gathering stage, summarized in Figure \ref{fig:dataset}, consisted of four steps:
\begin{compactitem}
    \item \textbf{Concept Generation} We used Gemini\footnote{Version used: \texttt{gemini-2.5-flash} (Released June 17, 2025). \label{fn:gemini}} to generate a large pool of concepts, which we manually curated to a final set of 138 coarse-grained, culturally-inclusive concepts (e.g., "train", "food"). Each concept is an abstract, semantic category that uses shared properties to group a broad, culturally-inclusive range of entities. The prompt for generating concepts can be found in Appendix \ref{apd:dataset_gathering}.
    \item \textbf{Concept De-duplication:} We use BGE-M3 \cite{chen2024bgem3embeddingmultilingualmultifunctionality} to de-duplicate concepts based on similarity with a threshold of 0.92.
    \item \textbf{Image Collection:} For each concept and a set of 16 diverse countries, we used the DuckDuckGo image search API \cite{deepanprabhu:duckduckgo} to retrieve the top 10 images using queries in both English (e.g., "train Japan") and the local native language.
    \item \textbf{Image De-duplication:} To ensure visual diversity, we performed two-stage de-duplication within each concept. First, near-exact duplicates were removed automatically using an embedding model. Subsequently, three human annotators, following the guidelines in Appendix \ref{apd:annotator_guide}, used a custom tool to manually filter out remaining images that depicted the same scene or object without meaningful variation in viewpoint or time of day.
\end{compactitem}
The final dataset contains 11,724 entries distributed across 138 concepts. Further statistics and samples are provided in Appendix \ref{appendix:data_stat} and \ref{apd:data_sample} respectively.

\subsubsection{RQ2:  Effect of Implicit Association Bias}
\label{sec:rq2}
\begin{figure}[htp]
  \centering
  \includegraphics[width=1.0\columnwidth]{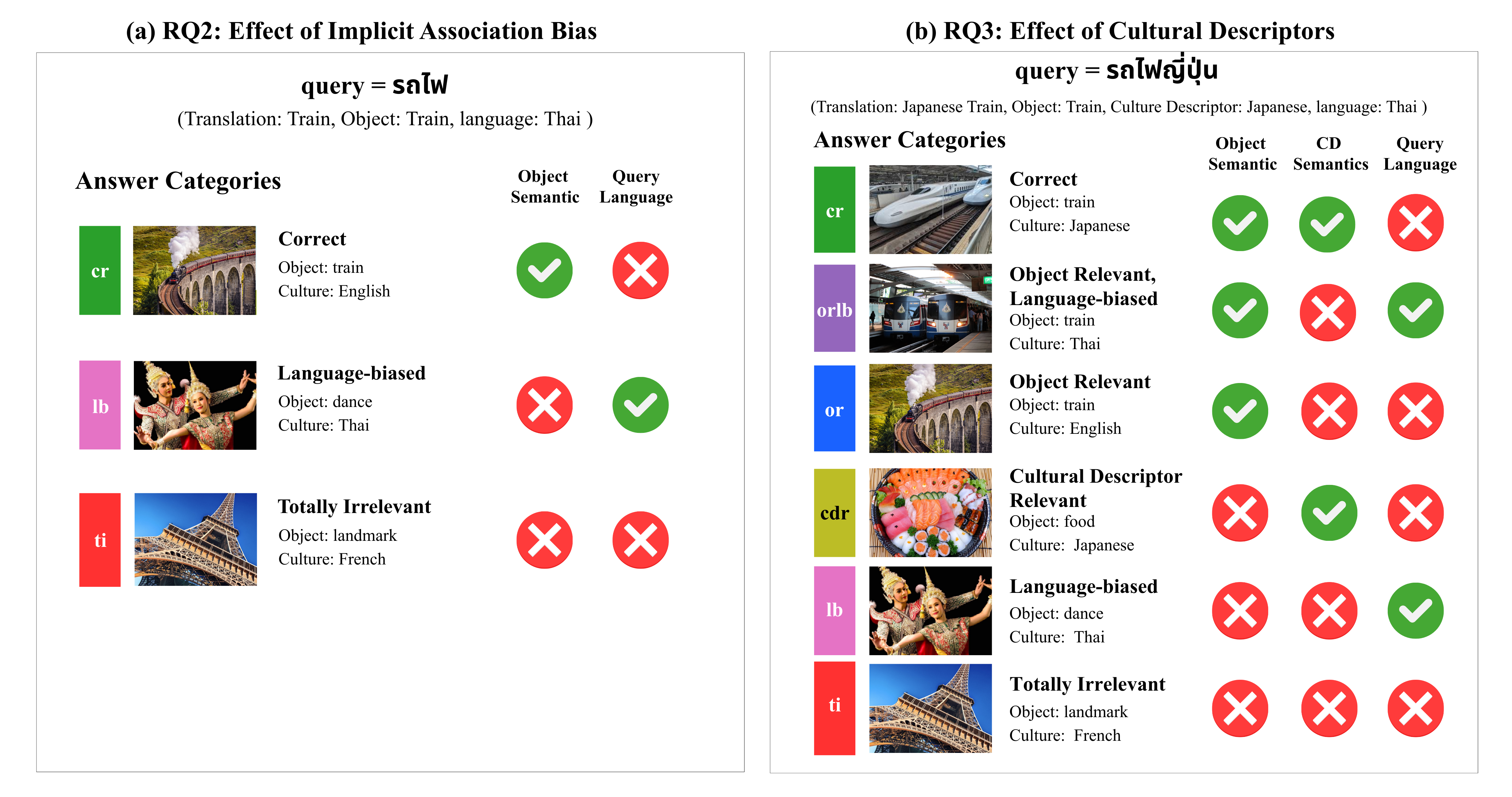}
  \caption{Illustration of \emph{association bias} evaluation. (a) RQ2 evaluates a Thai query for ``train'' against three candidates to distinguish semantic faithfulness from cultural relevance. (b) RQ3 uses the query ``train of Japan'' to measure the tension between explicit cultural descriptors and implicit language-cultural associations.}
  \label{fig:3query}
  \vspace{-0.5em}
\end{figure}

To evaluate implicit association bias,  we set up a forced-choice task designed to disambiguate between the model's reliance on semantic understanding (the concept) and its preference for cultural association. 
As illustrated in Figure~\ref{fig:3query}(a), for a given query (e.g., "train" in Thai), the model is presented with a triplet of image candidates representing three answer categories: 
(i) \textbf{Correct}: The model successfully identifies an image with a semantically relevant object (e.g., a train of any culture).
(ii) \textbf{Language-biased}: The model prioritizes culture-language association over semantic relevance (e.g., a Thai dance). 
(iii) \textbf{Totally irrelevant}: The model retrieves an image \emph{without} any semantic or cultural relevance (e.g., the Eiffel Tower). 

Note that we deliberately exclude images with relevance in terms of object semantics and query language to expose the prioritization of relevance.

With the constructed dataset, we can now measure the discrepancy between ideal retrieval outcomes and observed results, where bias arising from cultural association may intervene.
Ideally, the discrepancy should be zero when image retrieval depends solely on semantic relevance, and it should increase as the model’s preference tends towards images associated with the culture of the query, rather than semantic accuracy.
To quantify the discrepancy, we propose a metric called the \textbf{self-preference cultural bias score (SP)}, which can be computed as follows:
\vspace{-1em}

\begin{equation}
    M_k = \frac{1}{N} \sum_{i=1}^{N} \mathbb{I}\left( S_{k,i} = \max(S_{\text{cr},i}, S_{\text{lb},i}, S_{\text{ti},i}) \right) \label{eq:m_k_alt}
\end{equation}
\begin{equation}
    \text{SP} = \frac{M_{\text{lb}}}{M_{\text{cr}}} \label{eq:sp_score}
\end{equation}

where $M_k$ is the proportion of times a candidate of type $k$ receives the highest similarity score across $N$ total trials. The candidate type $k$ can be \textbf{Correct} (\texttt{cr}), \textbf{Language-biased} (\texttt{lb}), or \textbf{Totally Irrelevant} (\texttt{ti}). The similarity score for candidate type $k$ in trial $i$ is denoted by $S_{k,i}$. The indicator function $\mathbb{I}(\cdot)$ is 1 if the condition is true and 0 otherwise. The SP score (equation \ref{eq:sp_score}) is then the ratio of language-biased wins ($M_{\text{lb}}$) to correct answer wins ($M_{\text{cr}}$).
In this way, a higher SP score indicates stronger cultural self-preference over semantic faithfulness and thus a greater extent of association bias.

\subsubsection{RQ3: Effect of Cultural Descriptors}
\label{sec:rq3}
To investigate the model's adherence to explicit instructions against language-driven bias, we extend the forced-choice framework to include specific cultural cues.
We construct queries using an ``Object \& Cultural Descriptor'' format (e.g., ``Train of Japan'' formulated in Thai) to gauge the ``tugging effect'' between the explicit country mention and the implicit language-cultural association.
As illustrated in Figure~\ref{fig:3query}(b), we expand the answer category to six to isolate specific retrieval failures:
(i) \textbf{Correct}: The model successfully identifies an image with a semantically relevant object and cultural descriptor(e.g., a Shinkansen).
(ii) \textbf{Object Relevant, Language-biased}: The model successfully identifies an image with a semantically relevant object, but is distracted by the culture associated with the query language (e.g., a Thai train);
(iii) \textbf{Object Relevant}: The model successfully identifies an image with a semantically relevant object, but the culture is irrelevant to the query language and the cultural descriptor (e.g., British Train).
(iv) \textbf{Cultural Descriptor Relevant}: The model successfully identifies an image associated with the cultural descriptor, but misses the object's semantics (e.g., Sushi).
(v) \textbf{Language-biased}: The model fails to identify an image with a semantically relevant object but matches the query language's culture (e.g., Thai dance). 
(vi) \textbf{Totally Irrelevant}: The model captures neither the relevant object nor the culture associated with the descriptor or language. (e.g., Eiffel Tower).

To quantify the "tugging effect" between implicit language bias and explicit instructions, we calculate the \textbf{Similarity Drift} ($\Delta \text{sim}$). Given a base concept query $q_C$ and a culturally-augmented query $q_{C+CD}$, the drift for an image $I$ is defined as:
\begin{equation}
    \Delta \text{sim}(I) = S_{q_{C+CD}, I} - S_{q_C, I}
\end{equation}
A higher $\Delta \text{sim}$ for the Correct category relative to other categories with mismatch culture indicates a model's ability to prioritize explicit descriptors over inherent culture-language associations.

\section{Experimental Setup}
To answer our research questions, we conducted three main studies. 
\textbf{For RQ1}, we performed image-to-text retrieval on the Crossmodal-3600 dataset \cite{ThapliyalCrossmodal2022}.
The dataset offers a parallel multilingual text pool comprising native captions in 36 languages, making it suitable for auditing cross-lingual behavior in image–text retrieval, without implying any particular pattern of disparities. 
With this dataset, we compute Language-wise DLBKL by defining the prior $P(x)$ as a uniform distribution (see Appendix \ref{apd:country_dlbkl})".
We evaluate models using Accuracy@5, NDCG@10, LBKL@10, and our proposed DLBKL@10. 
\textbf{For RQ2}, we performed text-to-image retrieval on our newly created 3XCM benchmark, evaluating models using our proposed SP score.
\textbf{For RQ3}, we extend text-to-image retrieval of RQ2 by using the extended version of the 3XCM benchmark with six candidates and analyze the similarity drift.

We selected a representative suite of models spanning three distinct architectural paradigms: \textbf{Vision-Language Contrastive Models}: foundational models trained with a contrastive objective mainly on English data, \textbf{Cross-lingual Alignment Models}: models use knowledge distillation to explicitly align multilingual text encoders to a fixed, pre-trained vision space for multilingual capability, and \textbf{MLLM-Based Retrieval Embedders}: utilize MLLM as the backbone and finetune for a contrastive objective for retrieval. We refer readers to Appendix \ref{apd:full_model_name} for detailed architectural descriptions and specific model configurations.

\section{Experimental Results}
Our experiments are designed to provide empirical examinations of perspectival biases manifested in image-to-text and text-to-image retrievals.  

\subsection{Image-to-Text Evaluation (RQ1)}
All models exhibit some degree of linguistic prevalence bias, as shown in Table \ref{tab:rq1_result}.
This bias is most pronounced in the top-ranked results, as models tend to prioritize high-resource languages over others in the early ranks.
Appendix~\ref{apd:lang_rank_freq} provides visualizations revealing the dominance of high-resource languages in the top ranks and the overall disparity in retrieval frequency between language resource tiers.
Results for additional ranks and an example of a retrieval result can be found in Appendix~\ref{apd:example_rq1}.

\begin{table}[htbp]
    \centering
    \small
    \setlength{\tabcolsep}{3.5pt}
    \begin{tabular}{lcccc}
        \toprule
        \textbf{Model} & \textbf{Acc} & \textbf{LBKL} & \textbf{DLBKL} & \textbf{NDCG} \\
          & \textbf{@5$\uparrow$} & \textbf{@10$\downarrow$} & \textbf{@10$\downarrow$} & \textbf{@10$\uparrow$} \\
        \midrule
        \multicolumn{5}{l}{\textbf{Vision-Language Contrastive Models}}\\
        \hspace{10pt}CLIP-L/14 & 0.509 & 15.394 & 15.398 & 0.290 \\ 
        \hspace{10pt}CN-CLIP-L/14 & 0.355 & 15.287 & 15.291 & 0.207 \\ 
        \midrule
        \multicolumn{5}{l}{\textbf{Cross-lingual Alignment Models}}\\
        \hspace{10pt}XLM-R-L/14 & 0.924 & 12.775 & 12.793 & 0.736 \\ 
        \hspace{10pt}XLM-R-B/16plus & 0.968 & \textbf{12.651} & \textbf{12.669} & \textbf{0.791} \\ 
        \midrule
        \multicolumn{5}{l}{\textbf{MLLM-Based Retrieval Embedders}}\\
        \hspace{10pt}ColQwen2.5-3b-M & 0.894 & 13.654 & 13.667 & 0.605 \\ 
        \hspace{10pt}ColQwen2.5-7b-M & 0.926 & 13.439 & 13.454 & 0.665 \\ 
        \hspace{10pt}ColQwen2.5-v0.2 & 0.754 & 14.228 & 14.241 & 0.481 \\ 
        \hspace{10pt}GME-Qwen2-2B & 0.967 & 13.627 & 13.644 & 0.717 \\ 
        \hspace{10pt}GME-Qwen2-7B & \textbf{0.979} & 13.378 & 13.397 & 0.770 \\ 
        \hspace{10pt}Jina-E-v4 & 0.972 & 13.008 & 13.025 & 0.775 \\ 
        \bottomrule
    \end{tabular}
    \caption{Image-to-text retrieval on Crossmodal-3600. Bias is measured by LBKL and DLBKL, calculated in a language-wise manner. Explicit alignment models (XLM-R) show substantially lower bias.}
    \label{tab:rq1_result}
\end{table}

Crucially, the explicit alignment models (XLM-R series) achieve the lowest bias scores by a significant margin, with XLM-R-B/16plus demonstrating near-zero linguistic prevalence bias according to both metrics, while maintaining high retrieval accuracy. 
This provides strong initial evidence that direct alignment is a more effective strategy for enforcing language fairness than relying on emergent capabilities from large-scale pre-training.

Building on these observations, we note that LBKL and DLBKL \textbf{quantify distributional bias rather than relevance}, and therefore need \emph{not} correlate with accuracy or NDCG in Table \ref{tab:rq1_result}. To assess both correctness and fairness, these bias metrics should be interpreted jointly with accuracy (and/or NDCG). Finally, while LBKL/DLBKL capture cross-language imbalance, they do \emph{not} measure model self-preference (e.g., favoring the query language over others); we operationalize and evaluate that phenomenon with our SP score.

\subsection{Text-to-Image Evaluation (RQ2)}
Using the proposed 3XCM benchmark, we evaluated the association bias of several multimodal retrievers, ranging from CLIP to more recent models. In this evaluation, the semantic win rate ($M_{\text{cr}}$) serves as a proxy for raw performance, while the SP score quantifies cultural bias. 
We observe that the baseline CLIP and CN-CLIP models exhibit a significant cultural bias, \textbf{often preferring a culturally associated but semantically incorrect image}, as shown in Table \ref{tab:bias_scores}.
\begin{table}[htbp]
    \centering
    \small
    \setlength{\tabcolsep}{3.5pt}
    \begin{tabular}{lcccc}
    \toprule
    \textbf{Model} & \textbf{$M_{\text{cr}}\uparrow$} & \textbf{$M_{\text{lb}}\downarrow$} & \textbf{$M_{\text{ti}}\downarrow$} & \textbf{SP$\downarrow$} \\
    \midrule
    \multicolumn{5}{l}{\textbf{Vision-Language Contrastive Models}}\\
    \hspace{10pt}CLIP-L/14 & 51.24\% & 40.78\% & 7.98\% & 0.80\\
    \hspace{10pt}CN-CLIP-L/14 & 56.39\% & 31.65\% & 11.95\% & 0.56 \\
    \midrule
    \multicolumn{5}{l}{\textbf{Cross-lingual Alignment Models}}\\
    \hspace{10pt}XLM-R-L/14 & 85.53\% & 6.84\% & 7.63\% & 0.08\\
    \hspace{10pt}XLM-R-B/16plus & 87.54\% & \textbf{6.23\%} & 6.24\% & \textbf{0.07}\\
    \midrule
    \multicolumn{5}{l}{\textbf{MLLM-Based Retrieval Embedders}}\\
    \hspace{10pt}GME-Qwen2-2B & 83.34\% & 11.64\% & 5.02\% & 0.14 \\
    \hspace{10pt}GME-Qwen2-7B & 84.63\% & 11.26\% & \textbf{4.11\%} & 0.13 \\
    \hspace{10pt}ColQwen2.5-v0.2 & 82.10\% & 10.93\% & 6.97\% & 0.13 \\
    \hspace{10pt}ColQwen2.5-3B-M & 83.36\% & 10.65\% & 6.00\% & 0.13 \\
    \hspace{10pt}ColQwen2.5-7B-M & 84.07\% & 11.40\% & 4.53\% & 0.14 \\
    \hspace{10pt}Jina-E-v4 & \textbf{87.56\%} & 7.20\% & 5.24\% & 0.08 \\
    \bottomrule
    \vspace{-1.5em}
    \end{tabular}
    \caption{Results on the XCM benchmark for Self-Preference Cultural Bias. 
    }
    \label{tab:bias_scores}
    \vspace{-1em}
\end{table}

\begin{figure}[htbp]
  \centering
  \includegraphics[width=0.7\columnwidth]{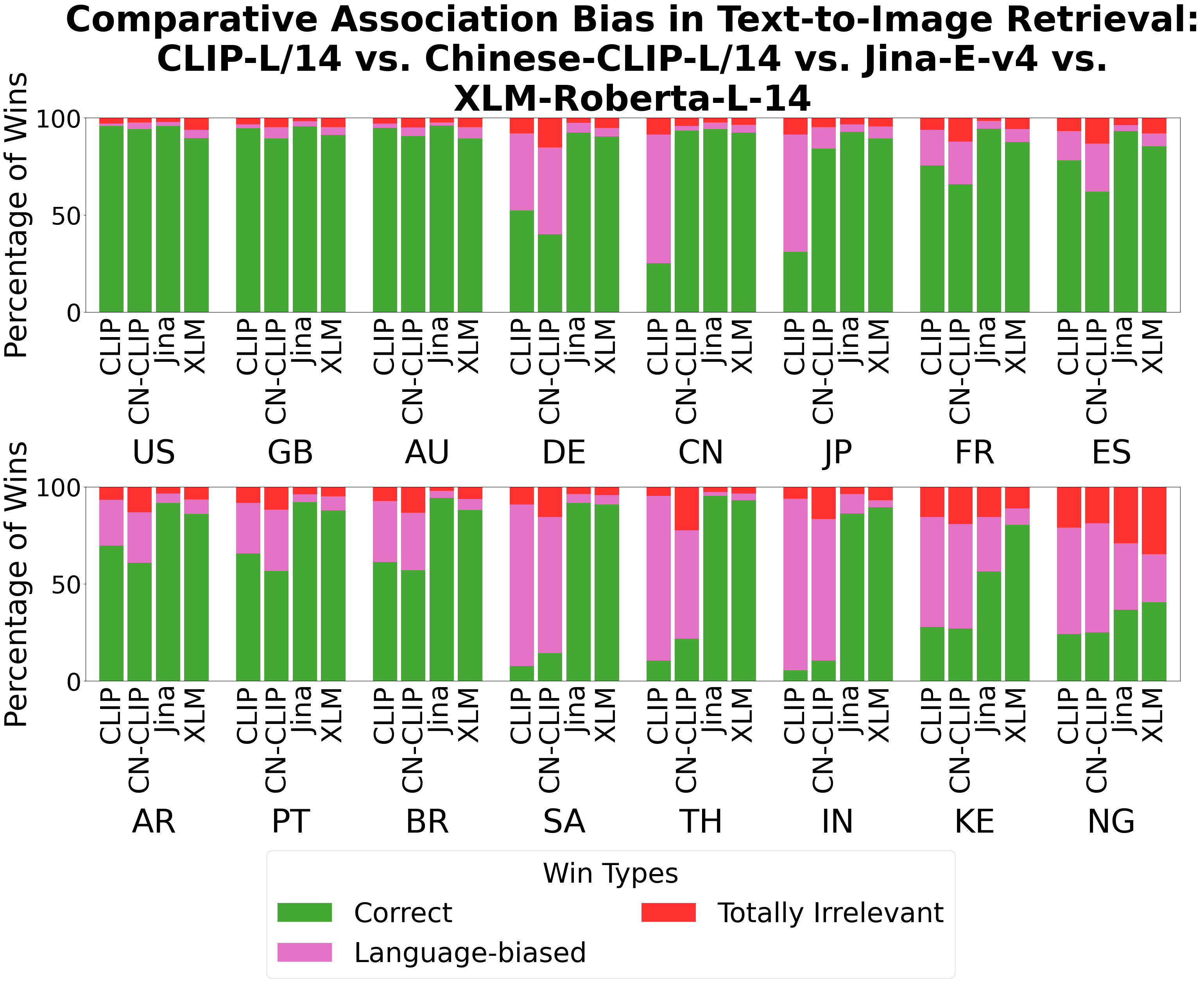}
  \caption{Association bias evaluation across four models reveals the limitations of monolingual training. The baseline CLIP shows significant cultural bias, which is exacerbated by region-specific fine-tuning as seen in CN-CLIP. In contrast, cross-lingual models like XLM-R and particularly Jina-E-v4 prove far more effective at overriding this bias and maintaining high semantic relevance across diverse countries.}
  \label{fig:self-preference-result-clip}
  \vspace{-0.5em}
\end{figure}

Our culture-specific analysis reveals that this self-preference is a symptom of missing linguistic knowledge, as shown in Figure~\ref{fig:self-preference-result-clip}. 
The CLIP-L/14 model, lacking a robust understanding of non-Latin scripts, defaults to matching cultural origin as a retrieval heuristic. 
Training on a large Chinese dataset (CN-CLIP) partially addresses this, improving performance for both Chinese and Japanese queries due to the shared logographic Kanji characters. 
However, this is a shallow fix that fails to generalize to other non-Latin scripts. 
In contrast, the text-aligned model (XLM-R-L/14) performs well across most languages, with a notable exception for queries in Yoruba (Nigeria). 
This challenge with low-resource languages persists even in more advanced architectures. For instance, MLLM-based models employ an LLM as their text encoder, leveraging its pre-training on web-scale multilingual data for a robust understanding of diverse languages. For the vision component, a Vision-Language Model (VLM) is used as the image encoder to improve contextual awareness. However, performance drops for low-resource languages.

\begin{figure*}[htbp]
  \centering
  \includegraphics[width=1.0\textwidth]{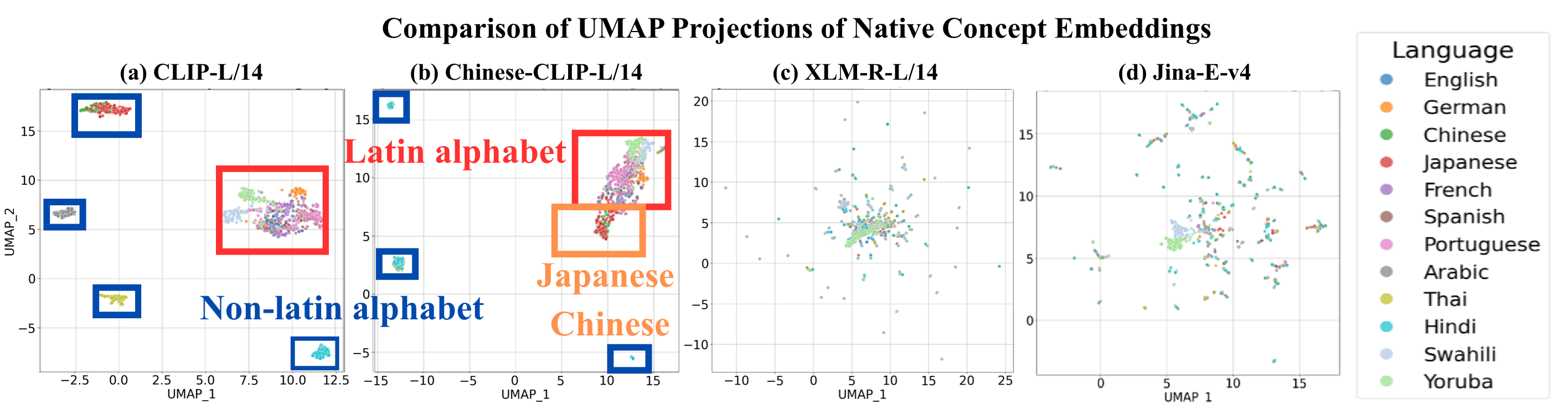}
  \caption{UMAP projection of native concept embeddings across four models: (a) CLIP-L/14 (non-Latin language separation), (b) CN-CLIP-L/14 (language family clustering), (c) XLM-R-L/14 (dense single-cluster unification), and (d) Jina-E-v4 (unified but dispersed cluster).}
  \label{fig:comparison_text_embedding}
\end{figure*}

This behavior is clearly visualized in the UMAP \cite{McInnes2018} projections of the text embeddings as shown in Figure \ref{fig:comparison_text_embedding}. 
The baseline CLIP-L/14 model exhibits a fractured embedding space, with non-Latin languages forming distinct clusters far from the main Latin-script cluster. 
This demonstrates a lack of shared semantic understanding. 
In the CN-CLIP model, the Chinese and Japanese embeddings shift closer to the Latin cluster, reflecting the targeted training, but other non-Latin languages remain isolated. 
In contrast, the explicit alignment model, XLM-R-L/14, successfully unifies the embedding space into a single, language-agnostic cluster, demonstrating a truly shared semantic representation across scripts. 
The only notable outlier is Yoruba, which was \emph{not} part of this specific model's alignment training. 
The MLLM model, Jina-E-v4, exhibits a similar but distinct pattern: it also forms a single, unified cluster, but the embeddings are more widely dispersed. This suggests a more flexible alignment that may capture finer semantic nuances between languages.

To validate these visual findings numerically, we calculated the silhouette score \cite{shahapure2020cluster} for each language's text embeddings. 
This analysis revealed a strong Pearson correlation (0.68) between a language's silhouette score and its measured SP score as shown in Appendix \ref{apd:sp_silhoutte}. 
This quantitatively reinforces that poor semantic understanding in the text encoder (as visualized by the disparate UMAP clusters) is a key driver of higher cultural association bias.

Both modern MLLM-based models and explicit alignment models drastically reduce the association bias compared to the baselines, achieving SP scores below 0.16. 
However, neither paradigm consistently outperforms the other on this specific task.

Model behavior remains consistent regardless of prompt phrasing, such as verbose descriptions (e.g., “a picture of a train for my homework”) or culture-agnostic instructions (e.g., “focus on semantics instead of textual language”) (Appendices \ref{apd:verbose-prompt-no-cd} and \ref{apd:system_prompt}).

\subsection{Cultural Descriptor Text-to-Image Evaluation (RQ3)}
\label{sec:rq3_result}
To investigate whether explicit instruction can override implicit association bias, we evaluate the models on the Extended 3XCM benchmark using queries with explicit Cultural Descriptors (CD) (e.g., ``\textit{Train of Japan}'' in Thai).

\paragraph{Adherence to Cultural Descriptors.}
Quantitative results demonstrate that most models are responsive to explicit cultural cues. As shown in Table~\ref{tab:bias_scores_6candidate}, \textbf{the correct answer} (matching both object and cultural descriptor) is the most frequently selected candidate across all high-performing models (XLM series, MLLM-based). 
This suggests that explicit cultural descriptors (e.g., the word ``Japan'') exert a stronger influence on the retrieval ranking than the implicit bias of the query language (e.g., Thai script), provided the model has sufficient multilingual capacity.

\begin{figure}[h]
  \centering
  \includegraphics[width=0.7\columnwidth]{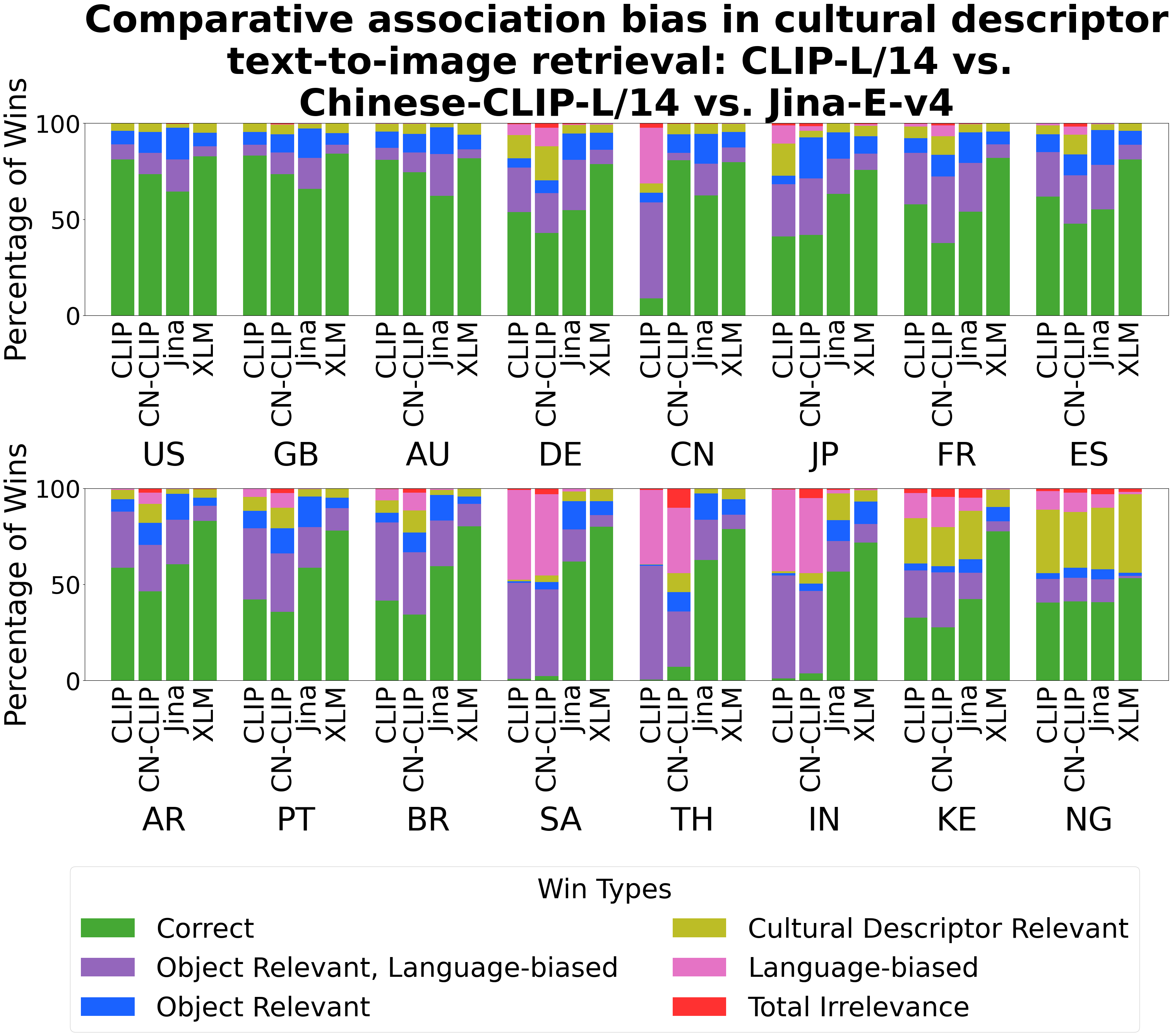}
  \caption{Comparative association bias across CLIP-L/14, CN-CLIP-L/14, Jina-embeddings-v4, and XLM-R-L/14 models in 16 countries. Stacked bars show win percentages for six categories, ranging from Correct to Total Irrelevance, highlighting performance variance across diverse cultural contexts.}
  \label{fig:self-preference-result-clip-rq3}
  \vspace{-1em}
\end{figure}

\begin{table}[htbp]
    \centering
    \small
    \setlength{\tabcolsep}{2.0pt} 
    \begin{tabular}{@{}lcccccc@{}}
    \toprule
    \textbf{Model} & \textbf{$M_{\text{cr}}$} & \textbf{$M_{\text{orlb}}$} & \textbf{$M_{\text{or}}$} & \textbf{$M_{\text{cdr}}$} & \textbf{$M_{\text{lb}}$} & \textbf{$M_{\text{ti}}$} \\
    & (\%) & (\%) & (\%) & (\%) & (\%) & (\%) \\
    \midrule
    \multicolumn{7}{l}{\textbf{Vision-Language Contrastive Models}}\\
    \hspace{5pt}CLIP-L/14 & 43.05 & 29.83 & 5.30 & 8.54 & 12.56 & 0.72 \\
    \hspace{5pt}CN-CLIP-L/14 & 41.97 & 24.54 & 9.81 & 10.04 & 11.21 & 2.42 \\
    \midrule
    \multicolumn{7}{l}{\textbf{Cross-lingual Alignment Models}}\\
    \hspace{5pt}XLM-R-L/14 & 77.83 & 7.21 & 6.85 & 7.55 & 0.26 & 0.30 \\
    \hspace{5pt}XLM-R-B/16plus & 72.98 & 9.88 & 9.70 & 6.57 & 0.40 & 0.48 \\
    \midrule
    \multicolumn{7}{l}{\textbf{MLLM-Based Retrieval Models}}\\
    \hspace{5pt}GME-Qwen2-2B & 60.01 & 18.38 & 11.49 & 8.05 & 1.56 & 0.50 \\
    \hspace{5pt}GME-Qwen2-7B & 59.75 & 20.84 & 10.79 & 6.81 & 1.38 & 0.43 \\
    \hspace{5pt}ColQwen2.5-v0.2 & 54.77 & 20.15 & 12.57 & 9.56 & 2.13 & 0.82 \\
    \hspace{5pt}ColQwen2.5-3B-M & 58.73 & 20.71 & 9.26 & 9.37 & 1.31 & 0.62 \\
    \hspace{5pt}ColQwen2.5-7B-M & 63.19 & 17.70 & 7.82 & 9.80 & 1.11 & 0.38 \\
    \hspace{5pt}Jina-E-v4 & 57.76 & 19.59 & 13.62 & 7.14 & 1.20 & 0.68 \\
    \bottomrule
    \end{tabular}
    \caption{Association bias results on the 3XCM benchmark (6 categories). $M_{\text{cr}}$, $M_{\text{orlb}}$, $M_{\text{or}}$, $M_{\text{cdr}}$, $M_{\text{lb}}$, and $M_{\text{ti}}$ represent Correct, Object Relevant with Language Bias, Object Relevant, Cultural Descriptor Relevant, Language-biased, and Totally Irrelevant, respectively.}
    \label{tab:bias_scores_6candidate}
    \vspace{-1.5em}
\end{table}

\paragraph{Failure Mode Analysis.}
Let us examine two failure modes presented by the error distribution in Figure \ref{fig:self-preference-result-clip-rq3}
First, when the model successfully retrieves the correct semantic object but fails to adhere to the explicit cultural descriptor (i.e., "Semantic Success, Cultural Failure"), we observe that the selection rate of the Object Relevant Language-biased category (which matches the object and the query language's culture, represented by purple color) is higher than that of the Object Relevant category (which matches the object with an unrelated culture, represented by blue color). This difference is significant ($p$-value < 0.05, Chi-squared test) for most models except the XLM family (see Appendix \ref{app:test_pairs}). The trend is most pronounced in non-Latin and low-resource languages, suggesting that when the explicit cultural constraint is violated, the model falls back on implicit language-driven associations.

Second, when the model fails to retrieve the correct semantic object (i.e., "Semantic Failure"), the model is more likely to select the Cultural Descriptor Relevant category (which matches the explicit cultural descriptor only, represented by yellow color) than the Language-biased category (which matches the query language's culture only, represented by pink color). This difference is significant ($p$-value < 0.05, Chi-squared test) for most models except the Vision-Language Contrastive group (CLIP and CN-CLIP), which lack robust non-Latin script understanding (see Appendix \ref{app:test_pairs}). This indicates that when the model fails to ground the visual object, it prioritizes the explicit cultural token (e.g., the country name) over the implicit association of the script.
(See Appendices~\ref{apd:sigtest} and \ref{apd:verbose-prompt-with-cd}).

\paragraph{Similarity Drift Analysis.}
The Similarity Drift ($\Delta \text{sim}$) metric measures the change in cosine similarity for a target image when a cultural descriptor is added to the query.
On average, adding a cultural descriptor successfully shifts the similarity distribution in the positive direction (toward the \textbf{Correct} category) across most models and languages. This confirms that the models generally recognize and attend to the explicit instruction.
However, a notable exception emerges. For the baseline CLIP model, queries in non-Latin scripts exhibit near-zero drift, indicating that the text encoder fails to process the cultural descriptor in these scripts, rendering the explicit instruction ineffective (see Appendix~\ref{apd:simdrift}).

\section{Discussion}
\noindent\textbf{Perspectival Biases and the Power of Explicit Alignment.} %
Our investigation establishes that perspectival bias manifests in two correlated forms: prevalence bias (RQ1) from distributional imbalances in training data, favoring high-resource languages, and association bias (RQ2) from spurious correlations between language scripts and cultural visual features. %
While both arise from training data and correlate empirically (high DLBKL models tend toward high SP), explicit cross-lingual alignment effectively reduces both, as seen in XLM-R models' low bias scores and high accuracy. %
This alignment not only mitigates prevalence and association biases but also enhances the model's ability to follow explicit cultural descriptors (CDs). %

\noindent\textbf{Guidance via Cultural Descriptors (CDs) and the Tugging Effect.}
CDs provide guidance to steer culture.
The positive Similarity Drift ($\Delta \text{sim}$) in RQ3 shows models can adhere to cultural descriptors, steering retrieval toward target cultures in most cases despite the tugging effect of query-language associations.
However, failure mode analysis reveals the tugging effect: when matching the object but not the CD, models revert to culture-language association over neutral alternatives.
This indicates association bias as a latent prior suppressed but not eliminated by guidance.
Modern MLLM-based models still exhibit issues with both biases and inconsistent compliance with CD guidance, especially for low-resource or non-Latin languages, where drifts are erratic or near-zero without robust multilingual support.

\section{Conclusion}

We distinguish prevalence bias, a distributional skew, from association bias, a semantic entanglement between language and culture. Even with explicit cultural descriptors, models exhibit a systematic ``tugging effect'' to language-culture association bias when semantic grounding succeeds but cultural disambiguation fails. These findings demonstrate that achieving cross-cultural equity requires more than naive scaling. Future work must prioritize training paradigms that actively de-correlate language from cultural features, enforce cultural constraints for steerability, or use explicit cross-lingual alignment. Only by pursuing these directions can we achieve a retrieval system that truly operates across global contexts with cultural controllability.

\section{Limitations}
Our work has several limitations. First, our \mbox{DLBKL} metric \textbf{measures fairness via distributional parity, \emph{not} semantic correctness.} It therefore \emph{cannot} distinguish between retrieving irrelevant documents and over-representing a language with relevant ones.
Second, the 3XCM benchmark simplifies culture by using country as a proxy, a necessary choice for tractability that does \emph{not} capture transnational or sub-national cultures. The benchmark's universal basic-level semantics (e.g., "food") and lack of accounting for polysemy also limit its representation of real-world query complexity, as its usage is intended to be used as a \textbf{challenging, controlled environment} used to uncover biases.

\bibliographystyle{unsrtnat}  
\bibliography{custom}

\appendix

\section{Comparison with Other Datasets}
\label{apd:compare_dataset}
We compare 3XCM with existing VQA benchmarks in Table~\ref{tab:benchmark_split}. The 3XCM dataset was developed to address the lack of cross-cultural parallelism in current resources.

\begin{table*}[!ht]
\centering
\footnotesize
\renewcommand{\arraystretch}{1.3}

\begin{subtable}{\textwidth}
\centering
\begin{tabular*}{\linewidth}{l @{\extracolsep{\fill}} ccc}
\toprule
\textbf{Feature} & \textbf{MaRVL} \cite{liu-etal-2021-visually} & \textbf{CulturalVQA} \cite{nayak-etal-2024-benchmarking} & \textbf{CVQA} \cite{romero2024cvqa} \\
\midrule
\textbf{Primary Task}   & Visual Reasoning & Visual QA & Visual QA \\
\textbf{Languages}      & 5 & 11 & 31 \\
\textbf{Granularity}    & Specific indigenous & Broad facets & Broad categories \\
\textbf{Parallelism}    & Low (cultural-specific) & Low (cultural-specific) & Low (region-specific) \\
\textbf{Bias Target}    & Reasoning Failure & Knowledge Gaps & Knowledge Gaps \\
\bottomrule
\end{tabular*}
\caption{Comparison of established cultural VQA benchmarks.}
\end{subtable}

\vspace{1.5em}

\begin{subtable}{\textwidth}
\centering
\begin{tabular}{lll}
\toprule
\textbf{Feature} & \textbf{RAVENEA} \cite{li2025ravenea} & \textbf{3XCM (Ours)} \\
\midrule
\textbf{Primary Task}   & VQA \& Image Cap. & \textbf{Cross-Modal Retrieval} \\
\textbf{Languages}      & 8 (Countries) & \textbf{16} \\
\textbf{Granularity}    & Varies (VQA:cultural-specific, IC: too broad) & \textbf{Universal basic-level} \\
\textbf{Parallelism}    & Low (Mixed) & \textbf{High (Cross-cultural)} \\
\textbf{Bias Target}    & Cultural Nuances & \textbf{Association Bias} \\
\bottomrule
\end{tabular}
\caption{Comparison with recent multi-task and cross-modal benchmarks.}
\end{subtable}

\caption{Comprehensive Benchmark Comparison.}
\label{tab:benchmark_split}
\end{table*}

\section{Language Resources}
\label{apd:language_resources}
To estimate language resource availability, we utilized the Distribution of Languages from the Common Crawl dataset (CC-MAIN-2025-18) \cite{CommonCrawlLanguages} as an approximation. Table \ref{tab:lang_dist_rq1} and \ref{tab:cc_lang_resources} present the resulting language composition for RQ1 and RQ2, respectively.

\begin{table}[!h]
    \centering
    \begin{tabular}{@{}llcr@{}}
        \toprule
        \textbf{Type} & \textbf{Language} & \textbf{ID} & \textbf{Distribution} \\
         &  &  & \textbf{(\%)} \\
        \midrule
        \multirow{3}{*}{High} & English & en & 43.9499 \\
                              & Russian & ru & 5.7614  \\
                              & German  & de & 5.5691  \\
        \midrule
        \multirow{12}{*}{Medium} & Japanese       & ja  & 4.9152 \\
                                 & Chinese-Simpl. & zh  & 4.8778 \\
                                 & Spanish        & es  & 4.5422 \\
                                 & French         & fr  & 4.3271 \\
                                 & Italian        & it  & 2.4060 \\
                                 & Portuguese     & pt  & 2.3369 \\
                                 & Polish         & pl  & 1.8744 \\
                                 & Dutch          & nl  & 1.8083 \\
                                 & Indonesian     & id  & 1.1759 \\
                                 & Turkish        & tr  & 1.1274 \\
                                 & Czech          & cs  & 1.0479 \\
                                 & Vietnamese     & vi  & 1.0213 \\
        \midrule
        \multirow{21}{*}{Low} & Korean         & ko  & 0.7865 \\
                              & Farsi          & fa  & 0.7087 \\
                              & Swedish        & sv  & 0.6736 \\
                              & Arabic         & ar  & 0.6722 \\
                              & Romanian       & ro  & 0.6374 \\
                              & Ukrainian      & uk  & 0.6079 \\
                              & Greek          & el  & 0.5651 \\
                              & Hungarian      & hu  & 0.5082 \\
                              & Danish         & da  & 0.4792 \\
                              & Thai           & th  & 0.4269 \\
                              & Finnish        & fi  & 0.3649 \\
                              & Norwegian      & no  & 0.3135 \\
                              & Hebrew         & he  & 0.2654 \\
                              & Croatian       & hr  & 0.2339 \\
                              & Hindi          & hi  & 0.2004 \\
                              & Bengali        & bn  & 0.1064 \\
                              & Telugu         & te  & 0.0213 \\
                              & Swahili        & sw  & 0.0102 \\
                              & Filipino       & fil & 0.0084 \\
                              & Maori          & mi  & 0.0014 \\
                              & Cusco Quechua  & quz & 0.0005 \\
        \bottomrule
    \end{tabular}
    \caption{Composition of Language Resources in the CommonCrawl Dataset (CC-MAIN-2025-18) for the language experimented in RQ1}
    \label{tab:lang_dist_rq1}
\end{table}

\begin{table}[h]
    \centering
    \footnotesize
    \begin{tabular}{@{}lllc@{}}
        \toprule
        \textbf{Code} & \textbf{Country} & \textbf{Language} & \textbf{Resources} \\
          & & & \textbf{(\%)} \\
        \midrule
        US  & America        & \multirow{3}{*}{English} & \multirow{3}{*}{43.950} \\
        GB  & Great Britain  &                          &                           \\
        AU  & Australia      &                          &                           \\
        \midrule
        DE  & Germany        & German                   & 5.569                     \\
        CN  & China          & Chinese                  & 4.878                     \\
        JP  & Japan          & Japanese                 & 4.915                     \\
        \midrule
        ES  & Spain          & \multirow{2}{*}{Spanish} & \multirow{2}{*}{4.542}    \\
        AR  & Argentina      &                          &                           \\
        \midrule
        FR  & France         & French                   & 4.327                     \\
        \midrule
        PT  & Portugal       & \multirow{2}{*}{Portuguese} & \multirow{2}{*}{2.337} \\
        BR  & Brazil         &                             &                        \\
        \midrule
        SA  & Saudi Arabia   & Arabic                   & 0.672                     \\
        TH  & Thailand       & Thai                     & 0.427                     \\
        IN  & India          & Hindi                    & 0.200                     \\
        KE  & Kenya          & Swahili                  & 0.010                     \\
        NG  & Nigeria        & Yoruba                   & 0.001                     \\
        \bottomrule
    \end{tabular}
    \caption{Composition of Language Resources in the CommonCrawl Dataset (CC-MAIN-2025-18) for the language experimented in RQ2} 
    \label{tab:cc_lang_resources}
\end{table}

\section{Evaluated Models and Details}
\label{apd:full_model_name}

In this work, we assess three distinct architectural paradigms for multimodal retrieval. The specific characteristics of these families are as follows:

\begin{itemize}
    \item \textbf{Vision-Language Contrastive Models:} These are foundational models trained primarily on English data. We include the original CLIP-L/14 as a powerful baseline, and CN-CLIP-L/14 to observe the effect of monolingual fine-tuning on a non-English corpus.

    \item \textbf{Cross-lingual Alignment Models:} These models use knowledge distillation to explicitly align multilingual text encoders to a fixed, pre-trained vision space for multilingual capability. We evaluate two variants of m-CLIP, which use XLM-RoBERTa as the text encoder (XLM-R-L/14 and XLM-R-B/16plus).

    \item \textbf{MLLM-Based Retrieval Embedders:} This modern paradigm adapts large, pre-trained Multimodal Language Models for retrieval. We evaluate several state-of-the-art models, including the ColQwen series (v0.2, 3b-M, 7b-M), GME models (Qwen2-2B, Qwen2-7B), and Jina-E-v4.
\end{itemize}

Table \ref{tab:model_aliases} provides the precise aliases, full model names, parameter counts, and citations used throughout the paper.

\begin{table*}[h]
\centering
\begin{tabular}{llc}
    \hline
    \textbf{Alias Used in Paper} & \textbf{Full Model Name} & \textbf{Parameter} \\
    \hline
    \multicolumn{3}{l}{\textbf{Vision-Language Contrastive Models}}\\
    \hspace{10pt}CLIP-L/14 & \verb|clip-vit-large-patch14|$^1$ & 427.6M \\
    \hspace{10pt}CN-CLIP-L/14 & \verb|Chinese-clip-vit-large-patch14|$^2$ & 406.2M \\
    \hline
    \multicolumn{3}{l}{\textbf{Cross-lingual Alignment Models}}\\
    \hspace{10pt}XLM-R-L/14 & \verb|XLM-Roberta-Large-Vit-L-14|$^6$ & 998.3M \\
    \hspace{10pt}XLM-R-L/16plus & \verb|XLM-Roberta-Large-Vit-B-16Plus|$^6$ & 768.9M \\
    \hline
    \multicolumn{3}{l}{\textbf{MLLM-Based Retrieval Embedders}}\\
    \hspace{10pt}ColQwen2.5-v0.2 & \verb|ColQwen2.5-v0.2|$^3$ & 3814.8M \\
    \hspace{10pt}ColQwen2.5-3B-M & \verb|ColQwen2.5-3b-multilingual-v1.0|$^3$ & 3994.6M \\
    \hspace{10pt}ColQwen2.5-7B-M & \verb|ColQwen2.5-7b-multilingual-v1.0|$^3$ & 8071.1M \\
    \hspace{10pt}GME-Qwen2-2B & \verb|gme-Qwen2-VL-2B-Instruct|$^4$ & 2209.0M \\
    \hspace{10pt}GME-Qwen2-7B & \verb|gme-Qwen2-VL-7B-Instruct|$^4$ & 7070.6M \\
    \hspace{10pt}Jina-E-v4 & \verb|jina-embeddings-v4|$^5$ & 3934.7M \\
    \hline
    \multicolumn{3}{l}{\parbox{\dimexpr\textwidth-2\tabcolsep\relax}{\vspace{2pt}\footnotesize The models are based on the following works: 1) \citet{DBLP:journals/corr/abs-2103-00020} for CLIP-L/14; 2) \citet{chinese-clip} for CN-CLIP-L/14; 3) \citet{faysse2025colpaliefficientdocumentretrieval} for ColQwen2 models; 4) \citet{zhang2025gmeimprovinguniversalmultimodal} for GME-Qwen2 models; 5) \citet{günther2025jinaembeddingsv4universalembeddingsmultimodal} for Jina-E-v4; and 6) \citet{carlsson-etal-2022-cross} for XLM-R-VL models.}} \\
    \hline
\end{tabular}
\caption{Aliases Used in Paper and Corresponding Full Model Names and Parameters}
\label{tab:model_aliases}
\end{table*}

\section{Details on LBKL and DLBKL}
\label{apd:dlbkl_details}

In Section~\ref{sec:rq1}, we utilize LBKL and DLBKL to measure prevalence bias. Here, we elaborate on the rationale behind these metrics.

In an ideal, bias-free retrieval scenario, the distribution of languages in the retrieved list should match the expected ''fair'' distribution (e.g., the ground truth distribution). LBKL measures the Kullback–Leibler divergence between these two distributions. In equation~\ref{lbkl equation}, $P(x)$ represents the proportion of a specific language in the ground truth, and $Q(x)$ represents the proportion in the retrieved set.

However, standard LBKL treats a deviation at rank 1 the same as a deviation at rank 100. In multimodal retrieval, users rarely scroll far, making top-rank bias significantly more harmful. Our proposed DLBKL in equation~\ref{dlbkl equation} addresses this by applying a discount factor derived from NDCG \citep{ndcg}. By weighting the observed proportion $Q(x)$ by the inverse log of the rank, we ensure that high-resource language dominance at the top of the list results in a significantly higher divergence score than if that same dominance occurred lower in the list.

\section{Formulation of Language-wise LBKL and DLBKL}
\label{apd:country_dlbkl}

In this section, we detail the calculation of two bias metrics: Language-wise LBKL (Country-LBKL) and Language-wise DLBKL (Country-DLBKL). 
Both metrics quantify the divergence between the retrieved country distribution and an ideal uniform distribution. 
The key distinction lies in how the predicted distribution is constructed: Country-LBKL treats the retrieved documents as an unweighted set, while Country-DLBKL applies rank-based discounting to prioritize higher-ranked results.

\subsection{Ground Truth Distribution (Shared)}
Let $C = \{c_1, c_2, ..., c_N\}$ be the set of target countries, where $N = |C|$. We define the ground truth distribution $P$ as:
\begin{equation}
    P(c_j) = \frac{1}{N}, \quad \forall c_j \in C
\end{equation}

\subsection{Country-LBKL (Unweighted)}
Country-LBKL evaluates bias based on the frequency of countries in the top-$k$ retrieved documents $D = [d_1, ..., d_k]$, treating all rank positions equally.

\subsubsection{Predicted Distribution (Unweighted)}
The unweighted proportion $Q(c_j)$ is calculated as:
\begin{equation}
    Q(c_j) = \frac{1}{k} \sum_{i=1}^{k} \mathbb{1}_{c_j}(d_i)
\end{equation}
where the indicator function $\mathbb{1}_{c_j}(d_i)$ is:
\begin{equation}
    \mathbb{1}_{c_j}(d_i) = 
    \begin{cases} 
        1 & \text{if } d_i \text{ is from } c_j \\
        0 & \text{otherwise}
    \end{cases}
\end{equation}

\subsubsection{Score Calculation}
The Country-LBKL score is the KL divergence between $P$ and $Q$. We simplify the substitution of $P(c_j)$ as follows:
\begin{equation}
\begin{split}
    \text{Country-LBKL} &= \sum_{c_j \in C} P(c_j) \log \left( \frac{P(c_j)}{Q(c_j)} \right) \\
    &= \frac{1}{N} \sum_{c_j \in C} \log \left( \frac{1}{N \cdot Q(c_j)} \right)
\end{split}
\end{equation}

\subsection{Country-DLBKL (Rank-Weighted)}
Country-DLBKL assigns higher importance to documents appearing earlier in the list using logarithmic discounting.

\subsubsection{Rank-Based Weighting}
For rank position $i$, the weight $w_i$ is:
\begin{equation}
    w_i = \frac{1}{\log_2(i + 1)}
\end{equation}
The total weight normalization factor is:
\begin{equation}
    W_{total} = \sum_{i=1}^{k} w_i = \sum_{i=1}^{k} \frac{1}{\log_2(i + 1)}
\end{equation}

\subsubsection{Predicted Distribution (Weighted)}
The weighted distribution $Q'(c_j)$ is calculated as:
\begin{equation}
    Q'(c_j) = \frac{\sum_{i=1}^{k} w_i \cdot \mathbb{1}_{c_j}(d_i)}{W_{total}}
\end{equation}

\subsubsection{Score Calculation}
The Country-DLBKL score is the KL divergence between $P$ and the weighted distribution $Q'$:
\begin{equation}
\begin{split}
    \text{Country-DLBKL} &= \sum_{c_j \in C} P(c_j) \log \left( \frac{P(c_j)}{Q'(c_j)} \right) \\
    &= \frac{1}{N} \sum_{c_j \in C} \log \left( \frac{1}{N \cdot Q'(c_j)} \right)
\end{split}
\end{equation}

\subsection{Implementation Note}
For both metrics, a small smoothing value is applied to $Q(c_j)$ and $Q'(c_j)$ in implementation to ensure numerical stability when a country is not represented in the retrieval results.

\section{Example of Prevalence Bias Evaluation}
\label{apd:example_rq1}
\begin{figure*}[t]
  \centering
\includegraphics[width=\textwidth, trim={2cm 11cm 2cm 2cm}, clip]{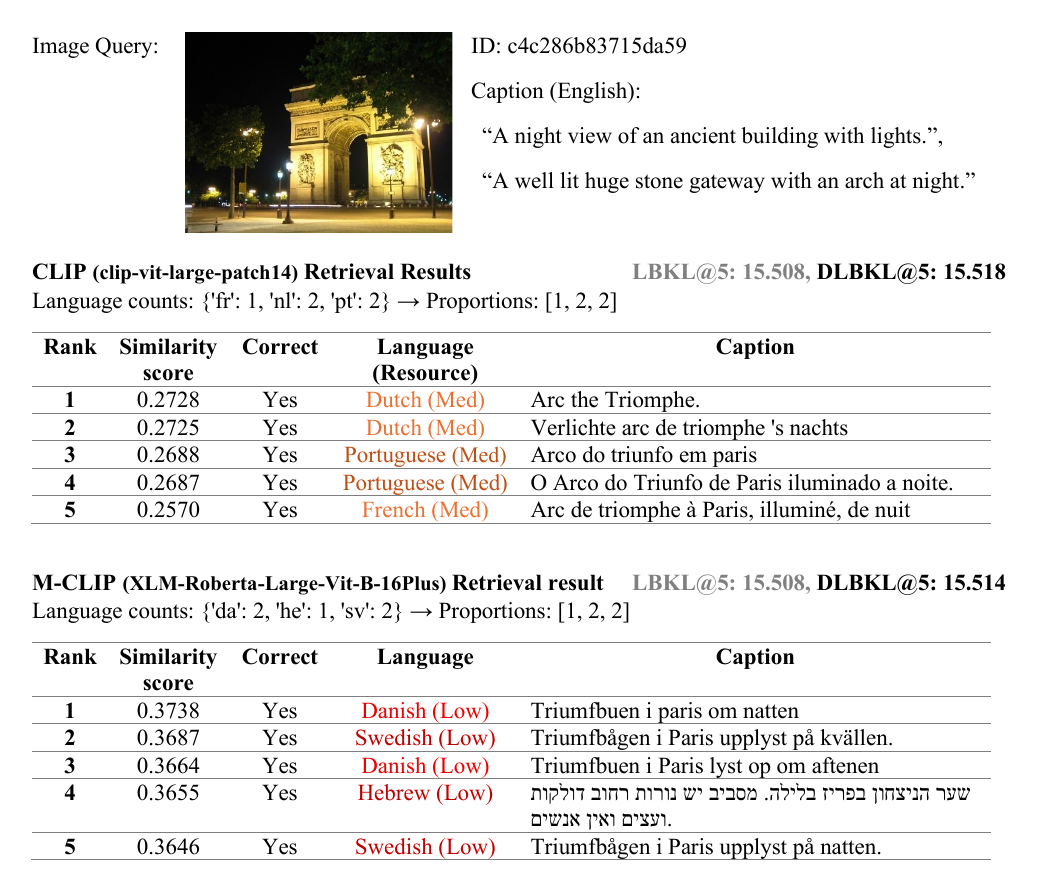}
    \caption{Qualitative Comparison of Image-to-Text Retrieval in Crossmodal-3600. Two retrieval results share identical semantic proportions and LBKL scores. Despite this semantic equality, the DLBKL score differs, capturing the model's implicit preference for specific linguistic groups in the ranking process.}
  \label{fig:cherry}
  \vspace{1.5em}
\end{figure*}

To further elaborate on the result of research question 1, we provide the example of retrieval from image to text from CLIP and M-CLIP in Figure \ref{fig:cherry}. 
We also provide the results of LBKL and DLBKL scores at other ranks in Table \ref{tab:rq1_result_appendix}.

\begin{table*}[!h]
    \centering
    \small
    \setlength{\tabcolsep}{4.5pt} 
    \begin{tabular}{lcccccccc}
        \toprule
        \textbf{Model} & \multicolumn{2}{c}{\textbf{@5}} & \multicolumn{2}{c}{\textbf{@25}} & \multicolumn{2}{c}{\textbf{@50}} & \multicolumn{2}{c}{\textbf{@99}} \\
        \cmidrule(lr){2-3} \cmidrule(lr){4-5} \cmidrule(lr){6-7} \cmidrule(lr){8-9}
        & \textbf{LBKL$\downarrow$} & \textbf{DLBKL$\downarrow$} & \textbf{LBKL$\downarrow$} & \textbf{DLBKL$\downarrow$} & \textbf{LBKL$\downarrow$} & \textbf{DLBKL$\downarrow$} & \textbf{LBKL$\downarrow$} & \textbf{DLBKL$\downarrow$} \\
        \midrule
        \multicolumn{9}{l}{\textbf{Vision-Language Contrastive Models}}\\
        \hspace{10pt}CLIP-L/14 & 15.846 & 15.848 & 14.706 & 14.711 & 14.158 & 14.162 & 13.629 & 13.631 \\
        \hspace{10pt}CN-CLIP-L/14 & 15.784 & 15.787 & 14.419 & 14.427 & 13.680 & 13.688 & 12.968 & 12.974 \\
        \midrule
        \multicolumn{9}{l}{\textbf{Cross-lingual Alignment Models}}\\
        \hspace{10pt}XLM-R-L/14 & 14.710 & 14.718 & 8.547 & 8.583 & 4.759 & 4.800 & 2.386 & 2.413 \\
        \hspace{10pt}XLM-R-B/16plus & 14.654 & 14.662 & 8.248 & 8.285 & 4.345 & 4.388 & 2.007 & 2.036 \\
        \midrule
        \multicolumn{9}{l}{\textbf{MLLM-Based Retrieval Embedders}}\\
        \hspace{10pt}ColQwen2.5-3B-M & 15.010 & 15.016 & 11.231 & 11.253 & 9.326 & 9.347 & 7.839 & 7.846 \\
        \hspace{10pt}ColQwen2.5-7B-M & 14.947 & 14.954 & 10.585 & 10.610 & 8.316 & 8.339 & 6.750 & 6.755 \\
        \hspace{10pt}ColQwen2.5-v0.2 & 15.340 & 15.346 & 12.176 & 12.199 & 10.561 & 10.586 & 9.315 & 9.331 \\
        \hspace{10pt}GME-Qwen2-2B & 15.144 & 15.151 & 10.457 & 10.494 & 7.664 & 7.710 & 5.322 & 5.368 \\
        \hspace{10pt}GME-Qwen2-7B & 15.035 & 15.042 & 9.766 & 9.805 & 6.503 & 6.552 & 4.024 & 4.068 \\
        \hspace{10pt}Jina-E-v4 & 14.781 & 14.789 & 9.320 & 9.354 & 6.079 & 6.120 & 3.734 & 3.769 \\
        \bottomrule
    \end{tabular}
    \caption{Image-to-text retrieval bias on Crossmodal-3600, measured by LBKL and DLBKL at various retrieval depths (k).}
    \label{tab:rq1_result_appendix}
\end{table*}

\section{Language and Rank Frequency Diagram}
\label{apd:lang_rank_freq}
\begin{figure*}[!ht]
  \centering
  \includegraphics[width=0.8\textwidth]{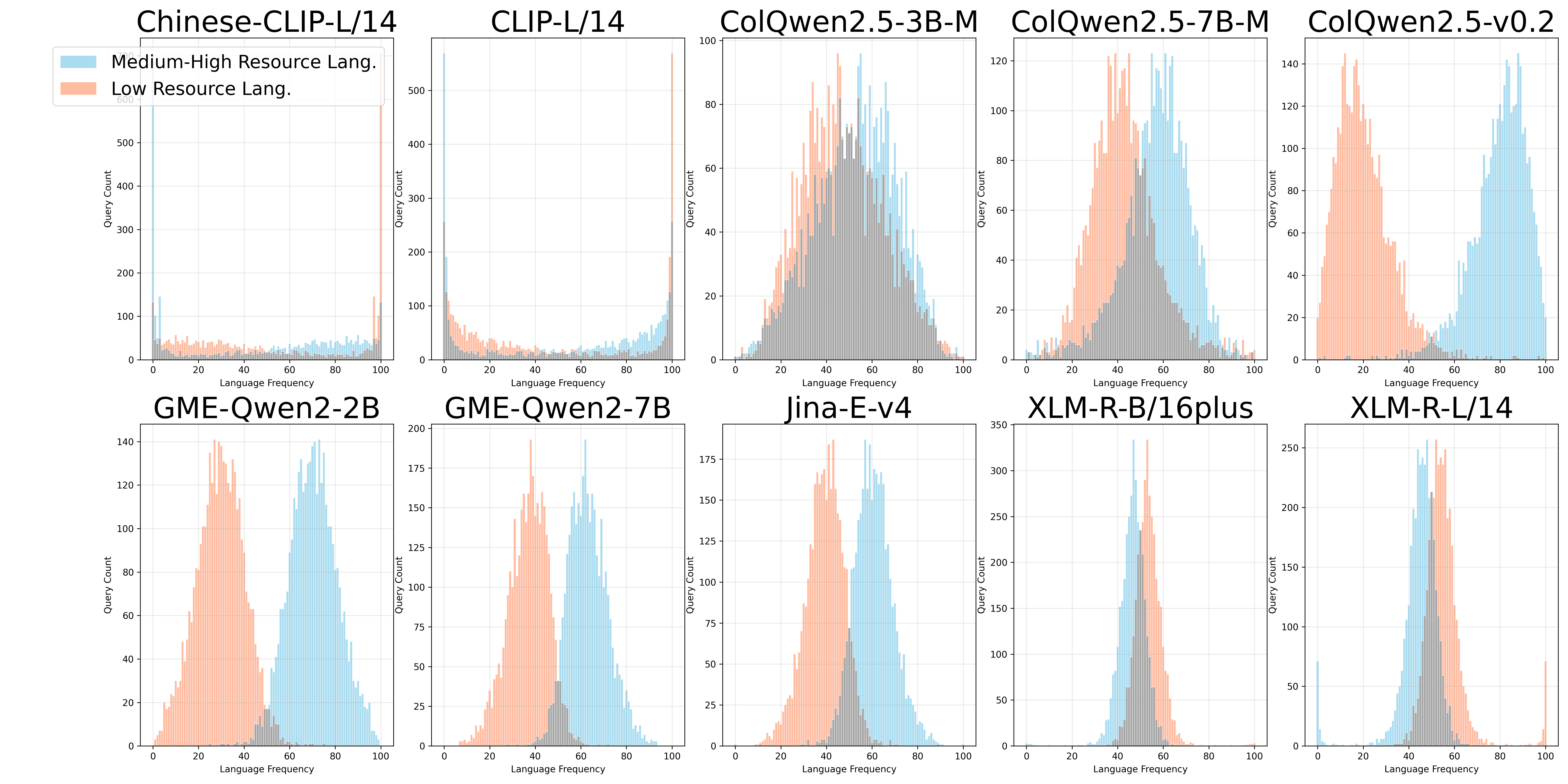}
  \caption{A language frequency histogram of each language group (see Appendix \ref{apd:language_resources}) for all models}
  \label{fig:langfreqhist_full}
\end{figure*}

\begin{figure*}[!ht]
    \centering
    \includegraphics[width=0.8\textwidth]{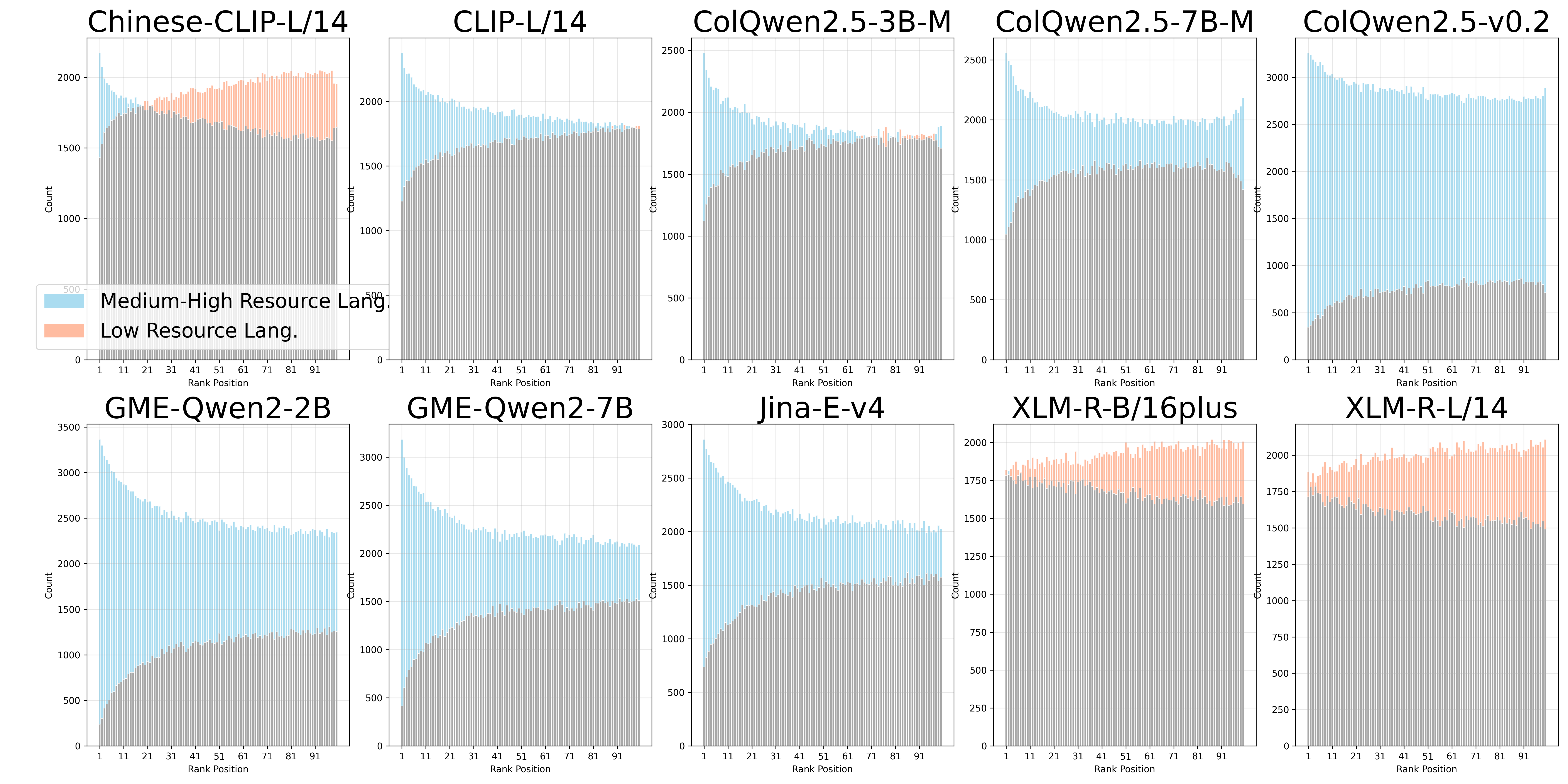}
  \caption{A frequency of language group (see Appendix \ref{apd:language_resources}) at each rank for all model}
  \label{fig:rankfreqhist_full}
\end{figure*}

To illustrate bias in image-to-text retrieval, we present a visualization of language groups categorized by resource level as shown in Appendix \ref{apd:language_resources}, showing both their overall retrieval frequency as shown in Figure \ref{fig:langfreqhist_full} and their frequency distribution across ranks as shown in Figure \ref{fig:rankfreqhist_full}.

\section{Culturally Relevant Concept Identification}
\label{apd:dataset_gathering}

To identify image concepts unique to each country, we first employed the Gemini\footref{fn:gemini}  as a tool for generating culturally relevant suggestions prior to data collection. The prompt used in this process is shown in Figure \ref{fig:app_generate_concept_prompt}.

\begin{figure}[H]
\centering
\begin{mdframed}[
  backgroundcolor=gray!10, 
  linewidth=1pt,           
  roundcorner=5pt,         
]
\fontsize{7.5pt}{9pt}\selectfont
\begin{verbatim}
list 100 concepts that unique and vary in these country
including China, India, Japan, saudi arabia, France,
German, Brazil, Kenya, Thailand, USA

like this

{"food":{"China":"Mala Xiang Guo", ..., "Thailand":
"Padthai", "USA":"hamburger"}, "costume":{"China":
"Hanfu", ..., "Thailand":"Sabai", "USA":"cowboy"}}
\end{verbatim}
\end{mdframed}
\caption{The prompt given to Gemini to generate unique country-specific image concepts.}
\label{fig:app_generate_concept_prompt}
\end{figure}

\section{Multi-Concept Detection in Images}
\label{apd:multi_concept_dectection}
To establish the self-preference cultural bias, the culturally relevant and non-relevant candidate images must \emph{not} share concepts with the text label. We utilized Gemini\footref{fn:gemini} with the following prompt in Figure \ref{fig:app_multiconcept_prompt} to identify all labels associated with an image.

\begin{figure}[H]
\centering
\begin{mdframed}[
  backgroundcolor=gray!10,
  linewidth=1pt,
  roundcorner=5pt,
]
\footnotesize 
Please classify the following image by assigning 
them to one or more of the following cultural 

categories:

\hspace{3ex} \{category\}.

\vspace{3ex} 

**Comprehensive Output Format (JSON):**
output as JSON for example:
\begin{verbatim}
{{
  "<INDEX>": ["<CATEGORY>", "<CATEGORY>", ...],
}}
\end{verbatim}
in the categories please order by priority (high to low).
\end{mdframed}
\caption{The prompt given to Gemini for multi-label image classification, where the \{category\} placeholder is dynamically populated with the full list of categories.}
\label{fig:app_multiconcept_prompt}
\end{figure}

\section{Annotator Guideline}
\label{apd:annotator_guide}
The guideline we provide to the annotators is to remove duplicates across multiple views. If an image depicts the same scene or object with no meaningful change, keep only one copy. Keep images if there is a significant variation. Allowed differences include time of day (e.g., day vs. night), viewpoint or angle (if the perspective changes enough that visual elements in the image are noticeably different). Minor or trivial variations are \emph{not} allowed as they would be too similar. This includes slight shifts, crops, or zooms of the same scene.

\section{Dataset Statistics}
\label{appendix:data_stat}
The distribution of cultural concepts in the 3XCM dataset is shown in Table \ref{tab:data_stat_concept}, with each concept being represented by approximately 85 images on average.
\begin{table}[ht]
    \centering
    \small 
    \begin{tabular}{@{}lc@{}} 
        \toprule
        \textbf{Country} & \textbf{Samples} \\
        \midrule
        Argentina    & 771 \\
        Australia    & 721 \\
        Brazil       & 724 \\
        China        & 727 \\
        France       & 760 \\
        Germany      & 744 \\
        India        & 774 \\
        Japan        & 944 \\
        Kenya        & 600 \\
        Nigeria      & 773 \\
        Portugal     & 824 \\
        Saudi Arabia & 619 \\
        Spain        & 841 \\
        Thailand     & 649 \\
        UK           & 644 \\
        USA          & 609 \\
        \midrule
        \textbf{Average} & \textbf{732.75} \\
        \bottomrule
    \end{tabular}
    \caption{Dataset Statistics of 3XCM per Country}
    \label{tab:dataset_stats_per_country}
\end{table}

\begin{table*}[!hp]
\centering
\label{table:concept_distribution}

\begin{tabular}{@{}ccc@{}} 
\hline
\begin{tabular}[t]{lr}
\multicolumn{2}{c}{\textbf{Concepts (A-G)}} \\
\hline
airlines & 24 \\
airport & 65 \\
alcohol drink & 26 \\
ancient city & 112 \\
ancient craft & 111 \\
ancient painting & 139 \\
animal & 100 \\
architecture & 107 \\
art & 89 \\
artwork & 26 \\
bag & 20 \\
bakery & 47 \\
banknotes & 69 \\
bathroom & 30 \\
bedroom & 77 \\
boat & 99 \\
bracelet & 62 \\
building & 196 \\
bus & 92 \\
bus station & 56 \\
capital & 147 \\
celebrity & 66 \\
child & 69 \\
\hline
\end{tabular}

&

\begin{tabular}[t]{lr}
\multicolumn{2}{c}{\textbf{Concepts (C-M)}} \\
\hline
cinema & 54 \\
coin & 182 \\
combination food & 72 \\
congress & 127 \\
costume & 119 \\
craft & 98 \\
dance & 145 \\
deep fried food & 30 \\
department store & 21 \\
dessert & 66 \\
devil & 43 \\
diningroom & 49 \\
doll & 131 \\
drink & 31 \\
dry heat food & 21 \\
embroidery style & 133 \\
fashion & 57 \\
festival & 145 \\
fire station & 162 \\
folk tale & 38 \\
folklore character & 90 \\
food & 52 \\
football player & 104 \\
\hline
\end{tabular}

&

\begin{tabular}[t]{lr}
\multicolumn{2}{c}{\textbf{Concepts (F-M)}} \\
\hline
formal uniform & 117 \\
fountain style & 91 \\
funeral & 141 \\
game & 76 \\
gas station & 63 \\
gathering place & 92 \\
ghost & 19 \\
graduated uniform & 78 \\
hat & 95 \\
headwear & 106 \\
historical event & 89 \\
historical figure & 81 \\
historical image & 120 \\
hot pot concept & 66 \\
hotel & 70 \\
house & 86 \\
instrument & 58 \\
lottery tickets & 48 \\
mailbox & 81 \\
major mountain range & 52 \\
major religious site & 97 \\
major river & 57 \\
map & 32 \\
market & 74 \\
\hline
\end{tabular}
\\
\hline
\begin{tabular}[t]{lr}
\multicolumn{2}{c}{\textbf{Concepts (M-P)}} \\
\hline
marriage ceremony & 95 \\
martial art & 96 \\
mask & 104 \\
military parade & 189 \\
moist heat food & 34 \\
museum & 99 \\
music band & 95 \\
mythical creature & 56 \\
mythological figure & 68 \\
native inhabitants & 163 \\
natural landmark & 152 \\
necklace & 63 \\
night view & 110 \\
older & 38 \\
painting & 128 \\
pants & 30 \\
people & 136 \\
poaching food & 25 \\
police station & 158 \\
popular street food & 98 \\
pottery style & 150 \\
priest & 68 \\
prime minister & 86 \\
\hline
\end{tabular}

&

\begin{tabular}[t]{lr}
\multicolumn{2}{c}{\textbf{Concepts (R-S)}} \\
\hline
religious building & 123 \\
restaurant & 38 \\
ritual & 108 \\
rural dwelling & 127 \\
sacred object & 101 \\
school & 120 \\
series & 40 \\
shirt & 64 \\
shopping mall & 91 \\
singer & 56 \\
snack & 56 \\
social custom & 147 \\
soldier & 167 \\
sport & 79 \\
stageplay & 108 \\
statue & 106 \\
street entertainment & 111 \\
street sign & 81 \\
street vendor cart & 32 \\
street view & 86 \\
symbolic bird & 84 \\
symbolic plant & 72 \\
\hline
\end{tabular}

&

\begin{tabular}[t]{lr}
\multicolumn{2}{c}{\textbf{Concepts (T-Z)}} \\
\hline
tattoo style & 59 \\
taxi & 102 \\
tea culture & 85 \\
textile pattern & 56 \\
tourist attraction & 130 \\
toy & 66 \\
train & 116 \\
train station & 128 \\
tree & 94 \\
tv program & 43 \\
unique art form & 119 \\
unique cuisine trait & 43 \\
unique food ingredient & 22 \\
unique natural phenomenon & 102 \\
unique transportation & 75 \\
university & 136 \\
wall painting & 65 \\
warrior & 79 \\
weapon & 75 \\
wedding & 108 \\
writing character & 15 \\
zoo & 79 \\
\multicolumn{2}{c}{}\\ 
\hline
\end{tabular}
\\
\end{tabular}
\caption{Distribution of Concepts and Image Counts}
\label{tab:data_stat_concept}
\end{table*}

\clearpage 

\section{3XCM Dataset Benchmark Samples}
\label{apd:data_sample}
\begin{figure}[h]
  \centering
  \includegraphics[width=0.6\textwidth]{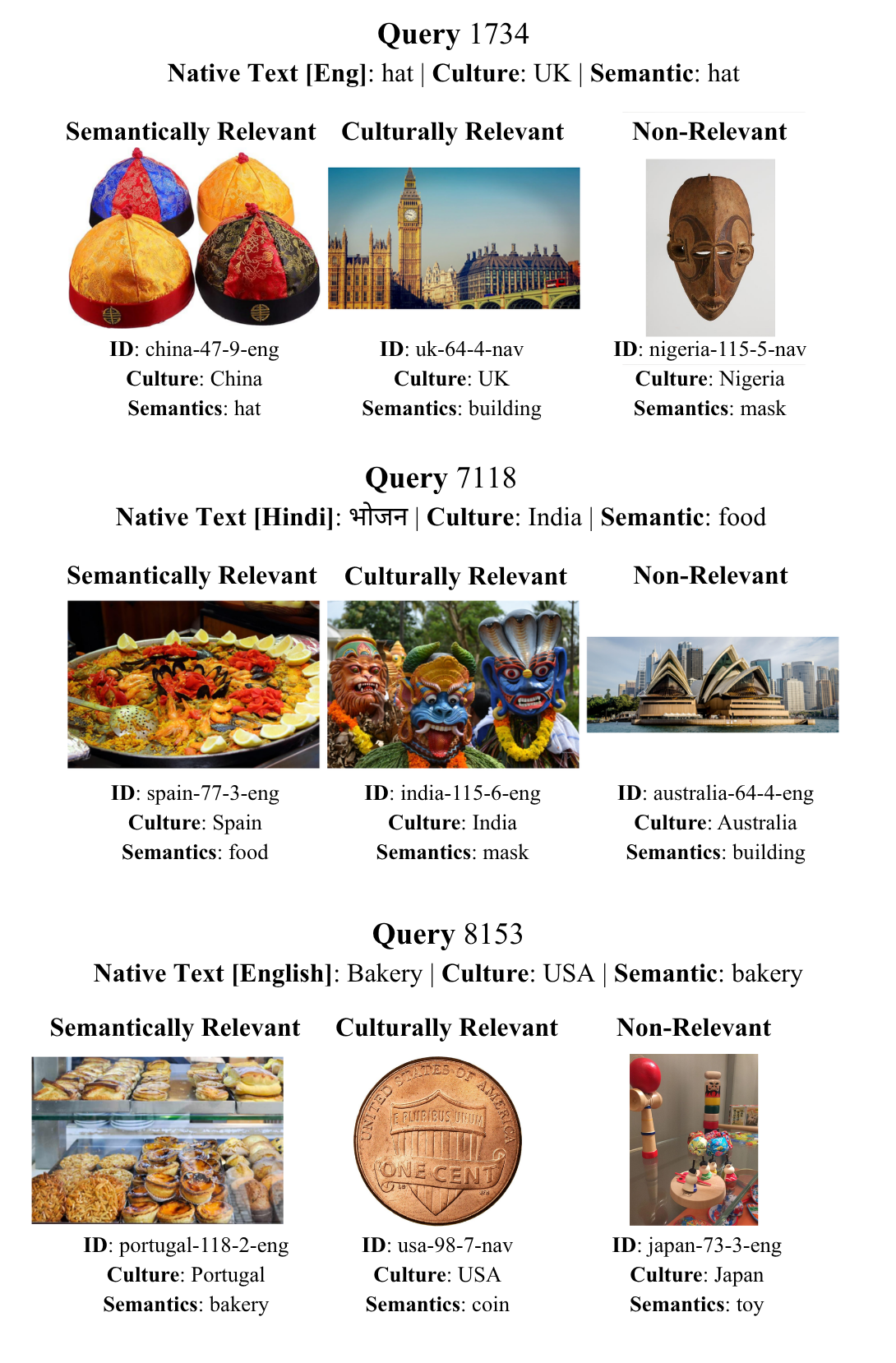}
  \caption{Benchmark examples from the 3XCM dataset featuring three retrieval candidates.}
  \label{fig:benchmark_samples}
\end{figure}

\begin{figure}[h]
  \centering
  \includegraphics[width=0.6\textwidth]{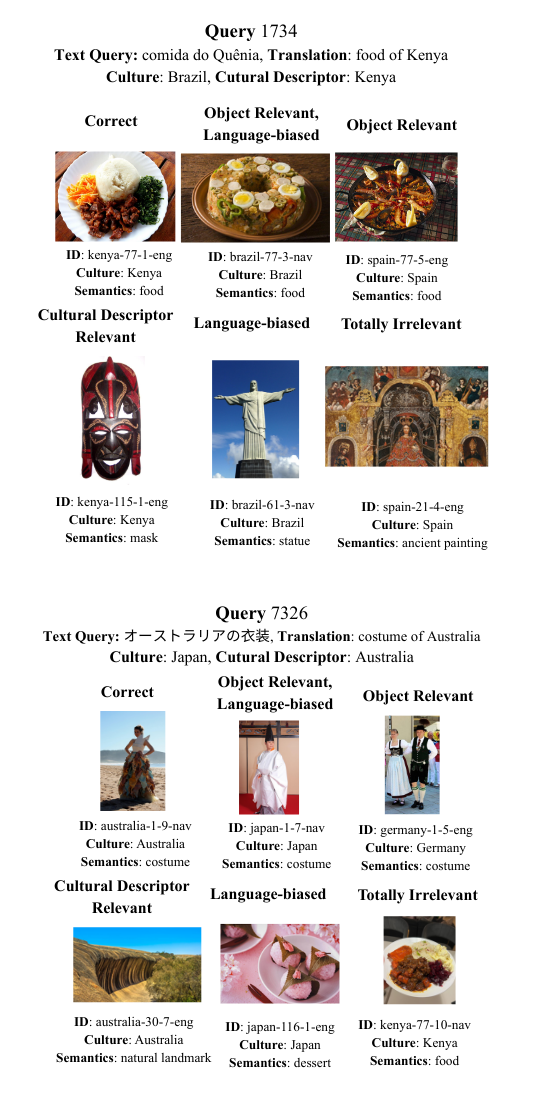}
  \caption{Benchmark examples from the 3XCM dataset featuring six retrieval candidates.}
  \label{fig:benchmark_samples_6}
\end{figure}

\begin{figure}[h]
\centering
\includegraphics[width=\textwidth]{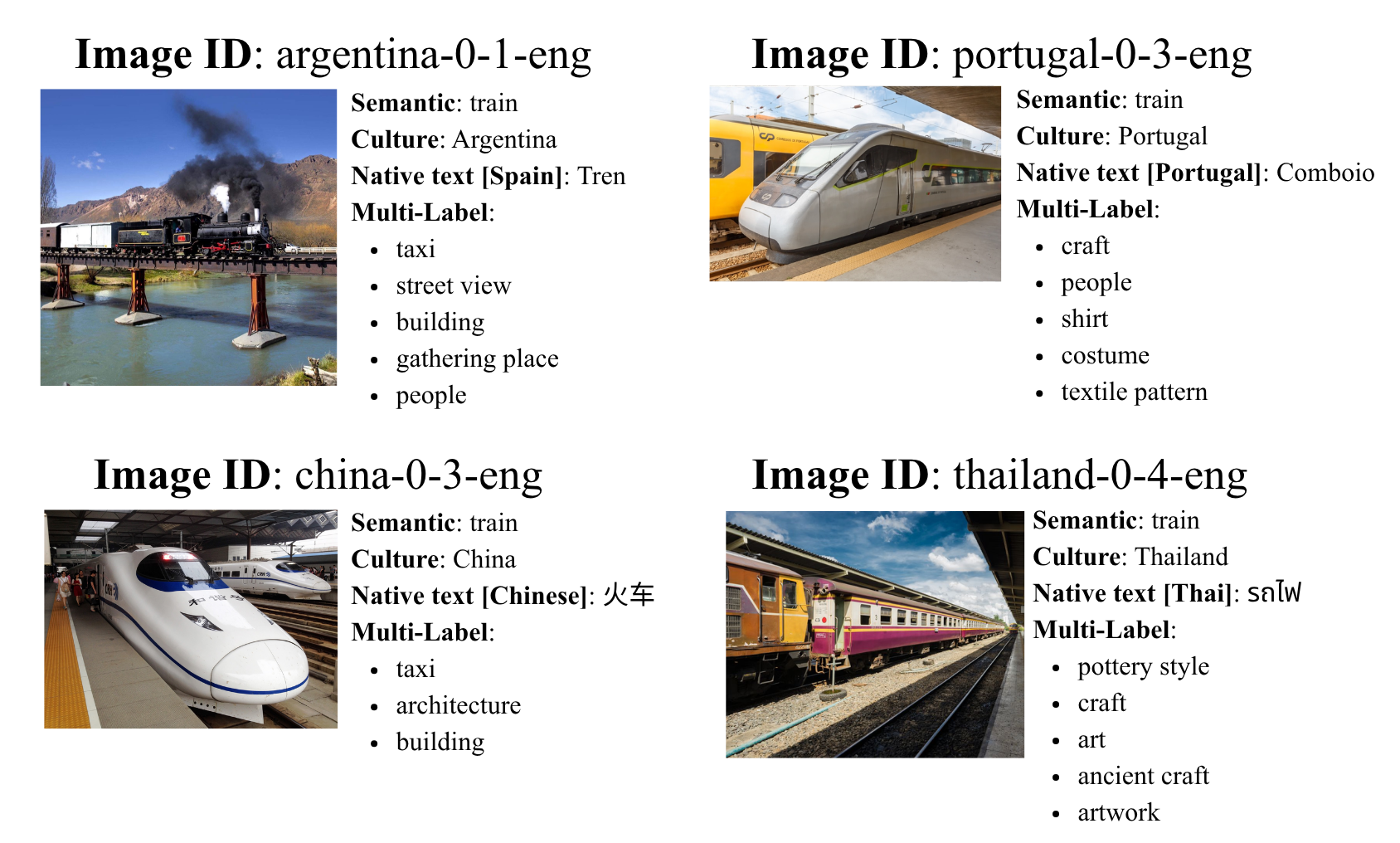}
  \caption{Metadata for image of 3XCM dataset benchmark}
  \label{fig:image_metadata}
\end{figure}

This research provides an association evaluation benchmark and image metadata. Examples of the benchmark for RQ2, RQ3, and image metadata are shown in Figure \ref{fig:benchmark_samples},  \ref{fig:benchmark_samples_6}, and \ref{fig:image_metadata} respectively.

\section{UMAP Analysis for Self-Preference Cultural Bias}
\label{apd:umap_anal}

\begin{figure*}[!hp]
    \centering 
    \includegraphics[width=0.8\columnwidth]{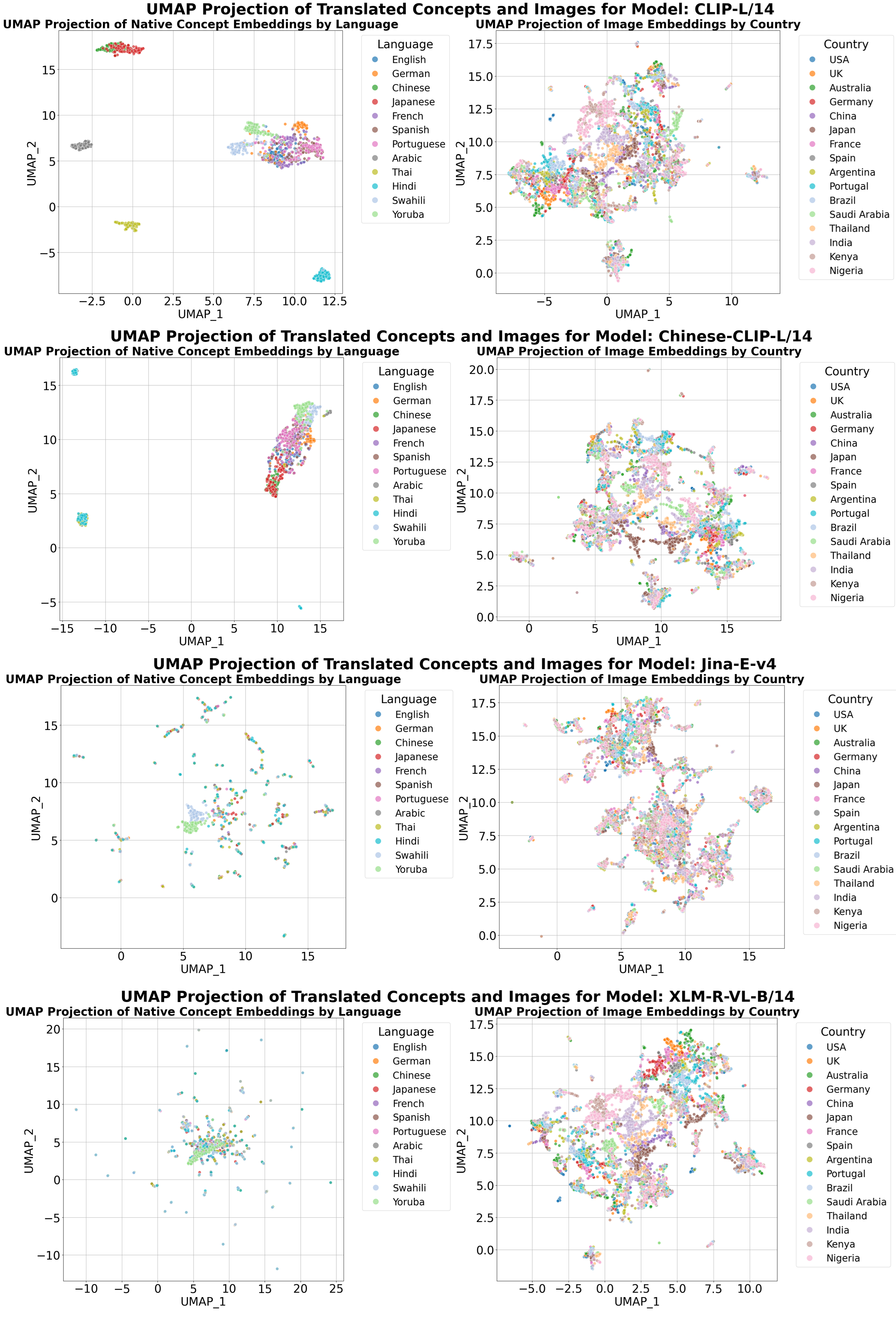} 
    \caption{The UMAP visualizations of the caption embeddings (left) and image embeddings (right) from the CLIP-L/14 model applied to our dataset.}
    \label{fig:apx_umap_part1}
\end{figure*}

\begin{figure*}[!h]
   \centering 
  \includegraphics[width=0.9\columnwidth]{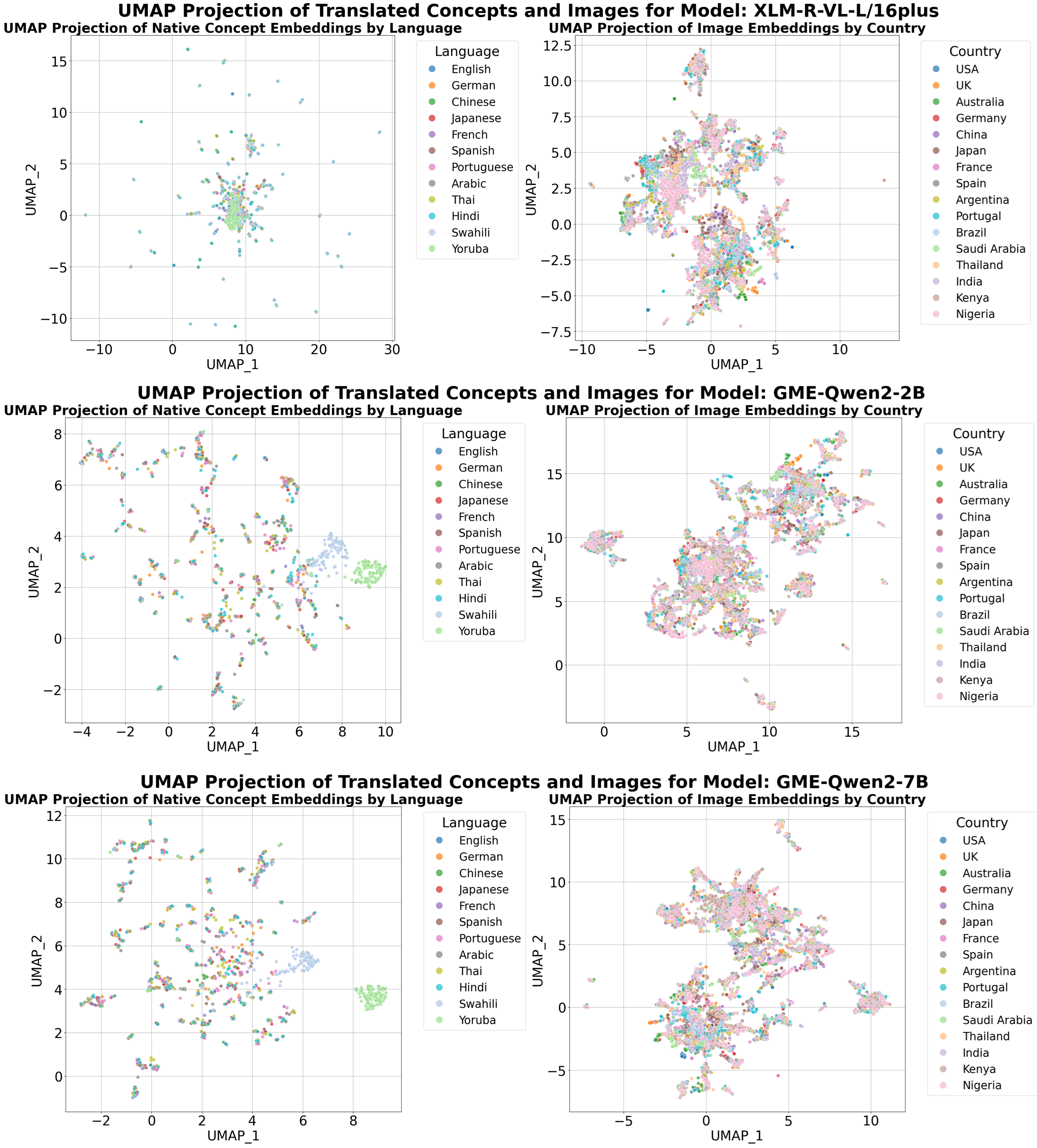}
  \caption{The UMAP visualizations of the caption embeddings (left) and image embeddings (right) from the CN-CLIP-L/14 model applied to our dataset.}
  \label{fig:apx_umap_part2}
\end{figure*}

To visualize cultural bias, the UMAP projections of text and image embeddings of all models as shown in Figure \ref{fig:apx_umap_part1} and \ref{fig:apx_umap_part2}. The text embeddings cluster strongly by language, a proximity that supersedes semantic content. Conversely, the image embeddings do \emph{not} exhibit strong country-based clustering, suggesting lower cultural bias. While other models show a similar, albeit less severe, tendency for text embeddings to be more biased than image embeddings, this effect is diminished in modern models. The GME-Qwen2 and Jina-E-v4 models only cluster very low-resource languages (Swahili, Yoruba), and the XLM-R models demonstrate superior alignment, forming a single central cluster. This discrepancy challenges retrieval systems: a query's text embedding is biased by its language, leading the system to favor images from the same cultural context over potentially more visually relevant content from others.

\clearpage
\clearpage

\section{Correlation Analysis for Self-Preference Cultural Bias}
\label{apd:sp_silhoutte}

\begin{figure}[t]
    \centering
    \includegraphics[width=\columnwidth]{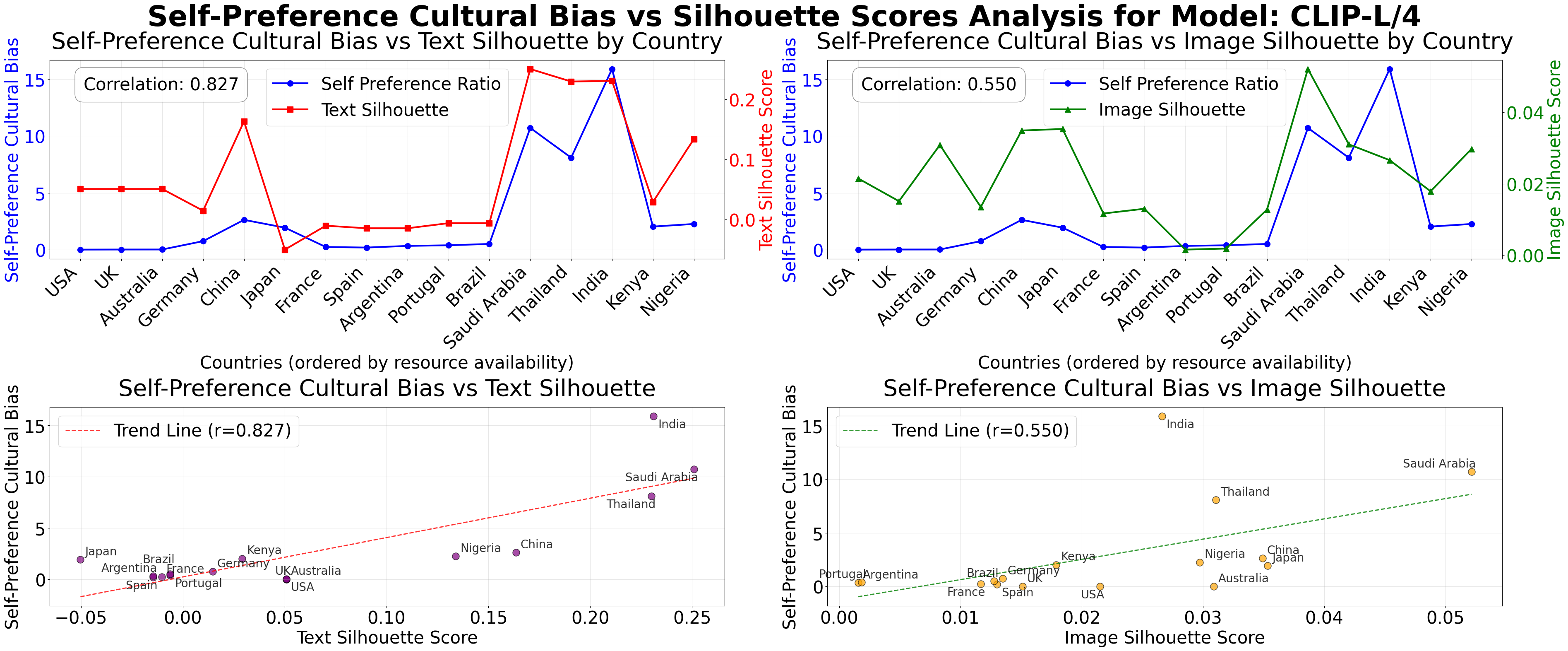}
    \caption{Comparison of Self-Preference Cultural Bias with Text and Image Silhouette}
    \label{fig:compare_sp_silhoette}
    \vspace{-1.5em}
\end{figure}

We investigate how unimodal bias, which our UMAP analysis shows is more severe in the text modality as shown in Appendix \ref{apd:umap_anal}, impacts cross-modal retrieval. To quantify this, we use the Silhouette score and find that high scores in low-resource languages correlate with the self-preference cultural bias score (SP), as shown in Table \ref{tab:silhouette_all_model}. We confirm this relationship by calculating the Pearson correlation between SP and the Silhouette scores. For example, CLIP-L/14's Text Silhouette score correlates strongly with SP (0.827), while its Image Silhouette correlation is only moderate (0.550), as shown in Figure \ref{fig:compare_sp_silhoette}. Across all tested models, the average correlations reveal that SP is predominantly driven by the text encoder, as shown in Table \ref{tab:compare_sp_silhoette}.

\begin{table}[h]
    \centering
    \begin{tabular}{lcc}
    \hline
    \textbf{Model} & \textbf{TSC} & \textbf{ISC} \\
    \hline
    CLIP-L/14 & 0.83 & 0.55 \\
    CN-CLIP-L/14 & -0.26 & 0.58\\
    XLM-R-VL-B/16 & 0.98 & -0.27\\
    XLM-R-VL-L/14 & 0.94 & 0.05\\
    Jina-E-v4 & 0.86 & 0.16 \\
    GME-Qwen2-2B & 0.80 & 0.09\\
    GME-Qwen2-7B & 0.64 & 0.07\\
    \hline
    \textbf{Average} & \textbf{0.68} & \textbf{0.18}\\
    \hline
    \end{tabular}
    \caption{This table presents a Pearson correlation analysis between model performance and bias. We measure the correlation between association and the quality of data clusters (via Silhouette Score) in both the text embedding space (TSC) and the image embedding space (ISC).}
    \label{tab:compare_sp_silhoette}
\end{table}

\begin{table*}[!h]
    \centering
    \footnotesize 
    \setlength{\tabcolsep}{1.5pt} 
    \begin{tabular}{@{}ll*{16}{c}@{}}
        \toprule
        \textbf{Model}& \textbf{Metrics} & \multicolumn{16}{c}{\textbf{Country}} \\
        \cmidrule(lr){3-18}
        & & \textbf{US} & \textbf{GB} & \textbf{AU} & \textbf{DE} & \textbf{CN} & \textbf{JP} & \textbf{FR} & \textbf{ES} & \textbf{AR} & \textbf{PT} & \textbf{BR} & \textbf{SA} & \textbf{TH} & \textbf{IN} & \textbf{KE} & \textbf{NG} \\
        \midrule
        
        \multirow{3}{*}{CLIP-L/14} 
        & SP $\downarrow$ & 0.01 & 0.02 & 0.02 & 0.76 & 2.63 & 1.94 & 0.24 & 0.19 & 0.34 & 0.39 & 0.51 & 10.71 & 8.09 & 15.88 & 2.04 & 2.27 \\
        & TS $\downarrow$ & 0.05 & 0.05 & 0.05 & 0.02 & 0.16 & -0.05 & -0.01 & -0.02 & -0.02 & -0.01 & -0.01 & 0.25 & 0.23 & 0.23 & 0.03 & 0.13 \\
        & IS $\downarrow$ & 0.02 & 0.02 & 0.03 & 0.01 & 0.03 & 0.04 & 0.01 & 0.01 & 0.00 & 0.00 & 0.01 & 0.05 & 0.03 & 0.03 & 0.02 & 0.03 \\
        \midrule
        
        \multirow{3}{*}{CN-CLIP-L/14} 
        & SP $\downarrow$ & 0.03 & 0.06 & 0.05 & 1.11 & 0.03 & 0.13 & 0.33 & 0.40 & 0.43 & 0.56 & 0.52 & 4.88 & 2.55 & 6.98 & 1.99 & 2.23 \\
        & TS $\downarrow$ & -0.05 & -0.05 & -0.05 & 0.00 & 0.13 & -0.07 & -0.03 & -0.03 & -0.03 & -0.01 & -0.01 & -0.40 & 0.93 & -0.37 & 0.00 & 0.03 \\
        & IS $\downarrow$ & 0.02 & 0.02 & 0.03 & 0.01 & 0.03 & 0.03 & 0.01 & 0.01 & 0.00 & 0.00 & 0.00 & 0.04 & 0.05 & 0.04 & 0.02 & 0.01 \\
        \midrule

        \multirow{3}{*}{Jina-E-v4} 
        & SP $\downarrow$ & 0.02 & 0.03 & 0.02 & 0.05 & 0.04 & 0.04 & 0.04 & 0.03 & 0.05 & 0.04 & 0.04 & 0.05 & 0.02 & 0.12 & 0.49 & 0.93 \\
        & TS $\downarrow$ & -0.01 & -0.01 & -0.01 & -0.02 & 0.01 & 0.00 & -0.01 & -0.02 & -0.02 & -0.02 & -0.02 & 0.02 & 0.01 & 0.00 & 0.00 & 0.13 \\
        & IS $\downarrow$ & -0.01 & 0.00 & -0.01 & -0.01 & 0.00 & 0.01 & 0.00 & 0.00 & -0.01 & -0.01 & -0.01 & 0.02 & 0.00 & 0.01 & 0.01 & 0.00 \\
        \midrule

        \multirow{3}{*}{XLM-R-L/14} 
        & SP $\downarrow$ & 0.05 & 0.04 & 0.07 & 0.05 & 0.04 & 0.07 & 0.08 & 0.08 & 0.09 & 0.08 & 0.07 & 0.05 & 0.04 & 0.04 & 0.11 & 0.61 \\
        & TS $\downarrow$ & -0.18 & -0.18 & -0.18 & -0.11 & -0.07 & -0.10 & -0.16 & -0.16 & -0.16 & -0.12 & -0.12 & -0.07 & -0.06 & -0.17 & -0.14 & 0.44 \\
        & IS $\downarrow$ & 0.01 & 0.02 & 0.03 & 0.01 & 0.03 & 0.03 & 0.01 & 0.01 & 0.01 & 0.01 & 0.01 & 0.06 & 0.04 & 0.03 & 0.02 & 0.03 \\
        \midrule

        \multirow{3}{*}{XLM-R-B/16plus} 
        & SP $\downarrow$ & 0.05 & 0.04 & 0.06 & 0.04 & 0.04 & 0.05 & 0.05 & 0.06 & 0.07 & 0.06 & 0.06 & 0.04 & 0.02 & 0.05 & 0.10 & 0.66 \\
        & TS $\downarrow$ & -0.24 & -0.24 & -0.24 & -0.21 & -0.19 & -0.19 & -0.24 & -0.23 & -0.23 & -0.23 & -0.23 & -0.19 & -0.19 & -0.23 & -0.23 & 0.49 \\
        & IS $\downarrow$ & 0.01 & 0.01 & 0.00 & 0.00 & 0.01 & 0.01 & 0.00 & 0.00 & -0.01 & 0.00 & -0.01 & 0.03 & 0.02 & 0.03 & 0.02 & 0.00 \\
        \midrule

        \multirow{3}{*}{GME-Qwen2-2B} 
        & SP $\downarrow$ & 0.02 & 0.03 & 0.02 & 0.10 & 0.04 & 0.04 & 0.05 & 0.05 & 0.08 & 0.08 & 0.09 & 0.14 & 0.10 & 0.24 & 1.24 & 2.02 \\
        & TS $\downarrow$ & 0.02 & 0.02 & 0.02 & 0.04 & 0.02 & 0.00 & 0.04 & 0.01 & 0.01 & 0.03 & 0.03 & 0.01 & 0.03 & 0.00 & 0.03 & 0.15 \\
        & IS $\downarrow$ & 0.00 & 0.00 & 0.01 & 0.00 & 0.02 & 0.01 & 0.00 & 0.00 & 0.00 & 0.00 & 0.00 & 0.03 & 0.02 & 0.02 & 0.01 & 0.01 \\
        \midrule

        \multirow{3}{*}{GME-Qwen2-7B} 
        & SP $\downarrow$ & 0.03 & 0.02 & 0.01 & 0.14 & 0.04 & 0.05 & 0.07 & 0.07 & 0.07 & 0.09 & 0.13 & 0.14 & 0.08 & 0.12 & 0.62 & 1.88 \\
        & TS $\downarrow$ & 0.04 & 0.04 & 0.04 & 0.05 & 0.02 & 0.02 & 0.07 & 0.02 & 0.02 & 0.04 & 0.04 & 0.09 & 0.11 & 0.03 & 0.04 & 0.14 \\
        & IS $\downarrow$ & 0.00 & 0.00 & 0.01 & 0.00 & 0.01 & 0.01 & 0.00 & 0.00 & 0.00 & 0.00 & 0.00 & 0.03 & 0.02 & 0.01 & 0.01 & 0.01 \\
        
        \bottomrule
    \end{tabular}
    \caption{Cross-Country and Cross-Model Comparison of Language and Cultural Bias Metrics. This table presents the results for the Self-Preference Cultural Bias score (SP), Text Silhouette (TS), and Image Silhouette (IS) scores across various multimodal retrievers for a selection of countries.}
    \label{tab:silhouette_all_model}
\end{table*}

\section{Detailed Results}
The full details of the RQ2 and RQ3 experiments, including all the win rates of all models, are illustrated in the Table \ref{tab:wins_all_model} and \ref{tab:wins_all_model_RQ3}, respectively.

\begin{table*}[!ht]
    \centering
    \footnotesize 
    \setlength{\tabcolsep}{1.5pt} 
    \begin{tabular}{@{}ll*{16}{c}@{}}
        \toprule
        \textbf{Model}& \textbf{Metrics} & \multicolumn{16}{c}{\textbf{Country}} \\
        \cmidrule(lr){3-18}
        & & \textbf{US} & \textbf{GB} & \textbf{AU} & \textbf{DE} & \textbf{CN} & \textbf{JP} & \textbf{FR} & \textbf{ES} & \textbf{AR} & \textbf{PT} & \textbf{BR} & \textbf{SA} & \textbf{TH} & \textbf{IN} & \textbf{KE} & \textbf{NG} \\
        \midrule
        
        \multirow{3}{*}{CLIP-L/14} 
        & $M_{\text{cr}} (\%)\uparrow$ & 95.73 & 94.57 & 94.73 & 52.42 & 25.17 & 31.07 & 75.53 & 78.00 & 69.65 & 65.78 & 61.33 & 7.75 & 10.48 & 5.56 & 27.83 & 24.19 \\
        & $M_{\text{lb}} (\%)\downarrow$ & 1.31 & 2.02 & 2.22 & 39.65 & 66.16 & 60.34 & 18.29 & 15.10 & 23.61 & 25.85 & 31.35 & 83.04 & 84.75 & 88.24 & 56.67 & 54.85 \\
        & $M_{\text{ti}} (\%)\downarrow$ & 2.96 & 3.42 & 3.05 & 7.93 & 8.67 & 8.59 & 6.18 & 6.90 & 6.74 & 8.37 & 7.32 & 9.21 & 4.78 & 6.20 & 15.50 & 20.96 \\
        \midrule
        
        \multirow{3}{*}{\shortstack{CN-CLIP\\-L/14}} 
        & $M_{\text{cr}} (\%)\uparrow$ & 94.25 & 89.44 & 90.57 & 40.05 & 93.40 & 84.20 & 65.79 & 61.95 & 60.83 & 56.55 & 57.04 & 14.38 & 21.88 & 10.47 & 27.00 & 25.10 \\
        & $M_{\text{lb}} (\%)\downarrow$ & 3.28 & 5.75 & 4.44 & 44.62 & 2.48 & 10.92 & 21.97 & 24.73 & 26.07 & 31.80 & 29.56 & 70.11 & 55.78 & 73.00 & 53.83 & 56.02 \\
        & $M_{\text{ti}} (\%)\downarrow$ & 2.46 & 4.81 & 4.99 & 15.32 & 4.13 & 4.88 & 12.24 & 13.32 & 13.10 & 11.65 & 13.40 & 15.51 & 22.34 & 16.54 & 19.17 & 18.89 \\
        \midrule

        \multirow{3}{*}{XLM-R-L/14} 
        & $M_{\text{cr}} (\%)\uparrow$ & 89.66 & 91.15 & 89.46 & 90.32 & 92.30 & 89.40 & 87.63 & 85.37 & 86.12 & 87.86 & 87.98 & 90.95 & 93.07 & 89.41 & 80.33 & 40.49 \\
        & $M_{\text{lb}} (\%)\downarrow$ & 4.11 & 4.04 & 5.83 & 4.44 & 4.13 & 6.26 & 6.58 & 6.54 & 7.39 & 7.16 & 5.80 & 4.85 & 3.54 & 3.75 & 8.50 & 24.71 \\
        & $M_{\text{ti}} (\%)\downarrow$ & 6.24 & 4.81 & 4.72 & 5.24 & 3.58 & 4.35 & 5.79 & 8.09 & 6.49 & 4.98 & 6.22 & 4.20 & 3.39 & 6.85 & 11.17 & 34.80 \\
        \midrule

        \multirow{3}{*}{\shortstack{XLM-R-B\\/16Plus}} 
        & $M_{\text{cr}} (\%)\uparrow$ & 91.95 & 91.30 & 90.71 & 93.95 & 94.50 & 91.62 & 90.66 & 88.47 & 87.81 & 90.05 & 89.36 & 92.57 & 95.22 & 91.09 & 83.17 & 40.88 \\
        & $M_{\text{lb}} (\%)\downarrow$ & 4.11 & 3.88 & 5.13 & 4.03 & 3.58 & 4.56 & 4.34 & 5.47 & 5.97 & 5.70 & 5.52 & 3.88 & 2.00 & 4.91 & 8.33 & 26.78 \\
        & $M_{\text{ti}} (\%)\downarrow$ & 3.94 & 4.81 & 4.16 & 2.02 & 1.93 & 3.82 & 5.00 & 6.06 & 6.23 & 4.25 & 5.11 & 3.55 & 2.77 & 4.01 & 8.50 & 32.34 \\
        \midrule

        \multirow{3}{*}{\shortstack{ColQwen2.5\\-3B-Multi}} 
        & $M_{\text{cr}} (\%)\uparrow$   & 95.24 & 93.79 & 95.01 & 83.20 & 94.91 & 90.99 & 89.61 & 89.06 & 87.94 & 89.44 & 89.09 & 87.40 & 90.60 & 81.65 & 45.67 & 27.30 \\
        & $M_{\text{lb}} (\%)\downarrow$ & 3.28  & 2.95  & 2.77  & 11.69 & 2.75  & 5.20  & 6.18  & 6.54  & 7.13  & 6.07  & 6.49  & 8.72  & 7.40  & 13.44 & 32.83 & 48.64 \\
        & $M_{\text{ti}} (\%)\downarrow$ & 1.48  & 3.26  & 2.22  & 5.11  & 2.34  & 3.82  & 4.21  & 4.40  & 4.93  & 4.49  & 4.42  & 3.88  & 2.00  & 4.91  & 21.50 & 24.06 \\
        \midrule

        \multirow{3}{*}{\shortstack{ColQwen2.5\\-7B-Multi}} 
        & $M_{\text{cr}} (\%)\uparrow$   & 96.55 & 96.74 & 96.81 & 84.27 & 95.32 & 91.73 & 90.13 & 88.35 & 84.82 & 88.71 & 88.67 & 86.75 & 91.99 & 86.82 & 42.17 & 32.08 \\
        & $M_{\text{lb}} (\%)\downarrow$ & 1.31  & 1.55  & 1.66  & 11.56 & 2.34  & 6.47  & 7.76  & 9.39  & 10.25 & 8.01  & 8.29  & 12.12 & 6.78  & 8.91  & 37.83 & 49.68 \\
        & $M_{\text{ti}} (\%)\downarrow$ & 2.13  & 1.71  & 1.53  & 4.17  & 2.34  & 1.80  & 2.11  & 2.26  & 4.93  & 3.28  & 3.04  & 1.13  & 1.23  & 4.26  & 20.00 & 18.24 \\
        \midrule

        \multirow{3}{*}{\shortstack{ColQwen2.5\\-v0.2}} 
        & $M_{\text{cr}} (\%)\uparrow$   & 93.10 & 89.60 & 92.09 & 83.60 & 94.09 & 88.55 & 90.79 & 87.51 & 86.64 & 90.29 & 90.61 & 85.14 & 88.60 & 79.59 & 38.83 & 29.88 \\
        & $M_{\text{lb}} (\%)\downarrow$ & 4.11  & 4.97  & 3.47  & 9.95  & 3.16  & 6.68  & 5.00  & 7.37  & 6.87  & 5.22  & 4.83  & 8.72  & 8.94  & 13.82 & 42.83 & 42.95 \\
        & $M_{\text{ti}} (\%)\downarrow$ & 2.79  & 5.43  & 4.44  & 6.45  & 2.75  & 4.77  & 4.21  & 5.11  & 6.49  & 4.49  & 4.56  & 6.14  & 2.47  & 6.59  & 18.33 & 27.17 \\
        \midrule

        \multirow{3}{*}{\shortstack{GME-Qwen2\\-2B-Instruct}} 
        & $M_{\text{cr}} (\%)\uparrow$ & 96.06 & 95.34 & 96.95 & 87.77 & 94.09 & 93.43 & 92.76 & 90.49 & 88.59 & 90.05 & 88.95 & 84.49 & 89.06 & 76.87 & 37.17 & 25.87 \\
        & $M_{\text{lb}} (\%)\downarrow$ & 2.30 & 2.48 & 1.66 & 8.47 & 3.58 & 3.92 & 4.74 & 4.64 & 7.26 & 7.04 & 7.87 & 11.63 & 8.63 & 18.48 & 46.00 & 52.26 \\
        & $M_{\text{ti}} (\%)\downarrow$ & 1.64 & 2.17 & 1.39 & 3.76 & 2.34 & 2.65 & 2.50 & 4.88 & 4.15 & 2.91 & 3.18 & 3.88 & 2.31 & 4.65 & 16.83 & 21.86 \\
        \midrule

        \multirow{3}{*}{\shortstack{GME-Qwen2\\-7B-Instruct}} 
        & $M_{\text{cr}} (\%)\uparrow$ & 95.40 & 95.34 & 96.53 & 84.68 & 95.05 & 92.47 & 90.39 & 90.73 & 90.27 & 88.35 & 85.50 & 85.46 & 90.91 & 85.92 & 55.17 & 29.62 \\
        & $M_{\text{lb}} (\%)\downarrow$ & 2.46 & 2.33 & 0.97 & 11.69 & 3.44 & 4.77 & 6.45 & 5.95 & 6.10 & 7.77 & 11.46 & 11.63 & 7.09 & 10.34 & 34.00 & 55.76 \\
        & $M_{\text{ti}} (\%)\downarrow$ & 2.13 & 2.33 & 2.50 & 3.63 & 1.51 & 2.76 & 3.16 & 3.33 & 3.63 & 3.88 & 3.04 & 2.91 & 2.00 & 3.75 & 10.83 & 14.62 \\

        \midrule
        \multirow{3}{*}{Jina-E-v4} 
        & $M_{\text{cr}} (\%)\uparrow$ & 95.89 & 95.65 & 95.98 & 92.47 & 94.22 & 92.79 & 94.34 & 93.22 & 91.70 & 91.99 & 94.20 & 91.60 & 95.38 & 86.30 & 56.50 & 36.74 \\
        & $M_{\text{lb}} (\%)\downarrow$& 1.97 & 2.48 & 1.66 & 4.97 & 3.44 & 3.92 & 4.08 & 3.09 & 4.80 & 4.13 & 3.73 & 4.68 & 2.00 & 9.95 & 27.83 & 34.15 \\
        & $M_{\text{ti}} (\%)\downarrow$ & 2.13 & 1.86 & 2.36 & 2.55 & 2.34 & 3.29 & 1.58 & 3.69 & 3.50 & 3.88 & 2.07 & 3.72 & 2.62 & 3.75 & 15.67 & 29.11 \\
        
        \bottomrule
    \end{tabular}
    \caption{Cross-Country and Cross-Model Comparison of Win Percentages. This table presents the results for the Correct, Language-biased, and Totally Irrelevant Win Percentages across various multimodal retrievers for a selection of countries.}
    \label{tab:wins_all_model}
\end{table*}

\begin{table*}[!htp]
    \centering
    \scriptsize 
    \setlength{\tabcolsep}{1.5pt} 
    \begin{tabular}{@{}ll*{16}{c}@{}}
        \toprule
        \textbf{Model}& \textbf{Metrics} & \multicolumn{16}{c}{\textbf{Country}} \\
        \cmidrule(lr){3-18}
        & & \textbf{US} & \textbf{GB} & \textbf{AU} & \textbf{DE} & \textbf{CN} & \textbf{JP} & \textbf{FR} & \textbf{ES} & \textbf{AR} & \textbf{PT} & \textbf{BR} & \textbf{SA} & \textbf{TH} & \textbf{IN} & \textbf{KE} & \textbf{NG} \\
        \midrule
        
        \multirow{6}{*}{CLIP-L/14} 
        & $M_{\text{cr}} (\%)$   & 80.39 & 82.17 & 79.94 & 43.36 & 4.12 & 35.23 & 50.33 & 48.28 & 40.86 & 37.14 & 36.80 & 0.81 & 0.76 & 1.03 & 13.79 & 17.16 \\
        & $M_{\text{orlb}} (\%)$ & 8.33 & 6.05 & 7.47 & 33.56 & 51.92 & 31.12 & 28.52 & 32.94 & 42.28 & 38.83 & 40.36 & 51.05 & 58.38 & 53.81 & 42.86 & 23.35 \\
        & $M_{\text{or}} (\%)$   & 7.68 & 8.06 & 8.71 & 3.62 & 3.71 & 4.64 & 8.28 & 11.30 & 8.04 & 8.86 & 7.52 & 1.13 & 0.76 & 0.26 & 2.49 & 5.03 \\
        & $M_{\text{cdr}} (\%)$  & 3.10 & 3.72 & 3.60 & 10.07 & 3.16 & 12.76 & 7.88 & 4.64 & 5.06 & 6.80 & 6.84 & 0.81 & 0.00 & 0.90 & 8.97 & 20.65 \\
        & $M_{\text{lb}} (\%)$   & 0.49 & 0.00 & 0.28 & 8.72 & 36.13 & 15.08 & 3.94 & 2.38 & 2.98 & 7.77 & 7.52 & 46.05 & 38.87 & 43.61 & 30.73 & 27.48 \\
        & $M_{\text{ti}} (\%)$   & 0.00 & 0.00 & 0.00 & 0.67 & 0.96 & 1.16 & 1.05 & 0.48 & 0.78 & 0.61 & 0.96 & 0.16 & 1.22 & 0.39 & 1.16 & 6.32 \\
        \midrule
        
        \multirow{6}{*}{\shortstack{CN-CLIP\\-L/14}} 
        & $M_{\text{cr}} (\%)$   & 62.91 & 64.65 & 63.21 & 22.82 & 78.43 & 25.00 & 25.10 & 26.99 & 26.72 & 22.45 & 22.71 & 0.97 & 12.35 & 5.55 & 14.12 & 19.35 \\
        & $M_{\text{orlb}} (\%)$ & 18.63 & 16.59 & 16.74 & 34.77 & 2.47 & 48.73 & 41.13 & 31.03 & 33.72 & 34.59 & 33.93 & 53.62 & 17.99 & 36.77 & 43.69 & 32.77 \\
        & $M_{\text{or}} (\%)$   & 14.38 & 12.71 & 11.89 & 7.38 & 14.29 & 14.98 & 10.12 & 12.84 & 10.51 & 12.74 & 12.04 & 1.29 & 14.63 & 5.55 & 4.65 & 5.68 \\
        & $M_{\text{cdr}} (\%)$  & 3.59 & 4.65 & 6.92 & 12.89 & 4.67 & 2.43 & 7.88 & 11.77 & 11.80 & 11.53 & 10.40 & 0.64 & 18.29 & 10.19 & 8.47 & 16.00 \\
        & $M_{\text{lb}} (\%)$   & 0.33 & 0.62 & 0.55 & 17.85 & 0.00 & 7.81 & 11.70 & 12.60 & 13.10 & 14.08 & 14.36 & 42.83 & 19.82 & 34.06 & 25.58 & 20.00 \\
        & $M_{\text{ti}} (\%)$   & 0.16 & 0.78 & 0.69 & 4.30 & 0.14 & 1.05 & 4.07 & 4.76 & 4.15 & 4.61 & 6.57 & 0.64 & 16.92 & 7.87 & 3.49 & 6.19 \\
        \midrule

        \multirow{6}{*}{XLM-R-L/14} 
        & $M_{\text{cr}} (\%)$   & 80.56 & 82.64 & 79.81 & 78.93 & 75.69 & 73.84 & 78.98 & 77.41 & 80.16 & 75.97 & 76.74 & 74.24 & 76.52 & 69.16 & 69.93 & 32.90 \\
        & $M_{\text{orlb}} (\%)$ & 8.17 & 6.05 & 6.50 & 9.53 & 9.48 & 8.76 & 8.28 & 10.34 & 9.73 & 11.89 & 13.27 & 7.25 & 8.54 & 9.03 & 4.65 & 6.19 \\
        & $M_{\text{or}} (\%)$   & 6.70 & 6.36 & 8.99 & 8.32 & 9.34 & 9.39 & 7.62 & 7.73 & 5.06 & 8.74 & 5.88 & 10.95 & 9.76 & 14.19 & 12.46 & 9.94 \\
        & $M_{\text{cdr}} (\%)$  & 4.58 & 4.96 & 4.01 & 2.55 & 4.67 & 7.49 & 4.60 & 4.40 & 4.93 & 3.40 & 3.56 & 6.44 & 4.88 & 5.29 & 9.47 & 33.29 \\
        & $M_{\text{lb}} (\%)$   & 0.00 & 0.00 & 0.55 & 0.54 & 0.41 & 0.32 & 0.53 & 0.00 & 0.13 & 0.00 & 0.14 & 0.00 & 0.30 & 0.77 & 1.00 & 7.10 \\
        & $M_{\text{ti}} (\%)$   & 0.00 & 0.00 & 0.14 & 0.13 & 0.41 & 0.21 & 0.00 & 0.12 & 0.00 & 0.00 & 0.41 & 1.13 & 0.00 & 1.55 & 2.49 & 10.58 \\
        \midrule

        \multirow{6}{*}{\shortstack{XLM-R-B\\/16Plus}} 
        & $M_{\text{cr}} (\%)$   & 77.29 & 78.45 & 75.24 & 72.89 & 71.98 & 71.84 & 73.19 & 72.29 & 74.19 & 68.57 & 73.46 & 68.28 & 69.51 & 64.00 & 64.45 & 32.77 \\
        & $M_{\text{orlb}} (\%)$ & 7.03 & 7.91 & 8.71 & 11.01 & 10.85 & 9.39 & 11.04 & 14.51 & 13.10 & 17.48 & 16.28 & 10.31 & 11.74 & 13.16 & 8.31 & 7.61 \\
        & $M_{\text{or}} (\%)$   & 12.25 & 9.46 & 12.03 & 12.62 & 12.23 & 12.03 & 9.99 & 9.51 & 7.52 & 10.19 & 7.80 & 12.88 & 14.33 & 14.97 & 13.79 & 9.16 \\
        & $M_{\text{cdr}} (\%)$  & 3.27 & 3.57 & 3.73 & 3.09 & 4.40 & 6.22 & 5.12 & 3.21 & 4.54 & 3.64 & 2.33 & 7.57 & 4.27 & 5.68 & 10.47 & 30.58 \\
        & $M_{\text{lb}} (\%)$   & 0.16 & 0.31 & 0.28 & 0.40 & 0.27 & 0.11 & 0.26 & 0.24 & 0.39 & 0.12 & 0.14 & 0.48 & 0.15 & 1.16 & 1.50 & 8.65 \\
        & $M_{\text{ti}} (\%)$   & 0.00 & 0.31 & 0.00 & 0.00 & 0.27 & 0.42 & 0.39 & 0.24 & 0.26 & 0.00 & 0.00 & 0.48 & 0.00 & 1.03 & 1.50 & 11.23 \\
        \midrule

        \multirow{6}{*}{\shortstack{ColQwen2.5\\-3B-Multi}} 
        & $M_{\text{cr}} (\%)$   & 70.42 & 72.25 & 70.95 & 54.23 & 69.51 & 57.81 & 54.27 & 56.48 & 54.22 & 54.25 & 55.95 & 62.96 & 49.85 & 39.74 & 26.08 & 23.35 \\
        & $M_{\text{orlb}} (\%)$ & 13.24 & 11.32 & 13.00 & 28.32 & 15.11 & 27.95 & 32.19 & 28.42 & 29.57 & 27.79 & 29.69 & 17.71 & 36.59 & 35.74 & 34.72 & 28.39 \\
        & $M_{\text{or}} (\%)$   & 15.03 & 13.02 & 14.11 & 14.77 & 11.40 & 11.71 & 10.78 & 13.20 & 13.10 & 15.66 & 11.35 & 14.01 & 9.45 & 15.10 & 6.98 & 3.35 \\
        & $M_{\text{cdr}} (\%)$  & 0.98 & 2.95 & 1.52 & 1.61 & 3.30 & 1.90 & 2.10 & 1.31 & 1.69 & 1.46 & 2.19 & 4.03 & 1.68 & 4.39 & 14.95 & 21.29 \\
        & $M_{\text{lb}} (\%)$   & 0.33 & 0.00 & 0.14 & 0.67 & 0.41 & 0.21 & 0.53 & 0.36 & 0.91 & 0.73 & 0.27 & 0.97 & 1.98 & 3.61 & 13.79 & 21.16 \\
        & $M_{\text{ti}} (\%)$   & 0.00 & 0.47 & 0.28 & 0.40 & 0.27 & 0.42 & 0.13 & 0.24 & 0.52 & 0.12 & 0.55 & 0.32 & 0.46 & 1.42 & 3.49 & 2.45 \\
        \midrule
        
         \multirow{6}{*}{\shortstack{ColQwen2.5\\-7B-Multi}} 
        & $M_{\text{cr}} (\%)$   & 69.44 & 71.94 & 69.71 & 49.26 & 64.56 & 56.65 & 55.19 & 49.11 & 51.62 & 54.37 & 56.09 & 63.45 & 49.09 & 49.94 & 31.56 & 27.10 \\
        & $M_{\text{orlb}} (\%)$ & 14.05 & 11.47 & 11.20 & 38.93 & 20.05 & 30.49 & 34.30 & 35.67 & 34.50 & 29.85 & 33.38 & 19.81 & 39.18 & 28.13 & 27.08 & 22.06 \\
        & $M_{\text{or}} (\%)$   & 14.71 & 13.33 & 16.32 & 9.93 & 11.68 & 10.86 & 7.62 & 9.99 & 10.51 & 12.14 & 8.21 & 9.82 & 6.86 & 13.55 & 6.98 & 5.29 \\
        & $M_{\text{cdr}} (\%)$  & 1.63 & 2.79 & 2.63 & 1.21 & 3.43 & 1.90 & 2.50 & 3.45 & 2.08 & 2.79 & 2.05 & 4.99 & 2.29 & 5.55 & 18.60 & 24.77 \\
        & $M_{\text{lb}} (\%)$   & 0.00 & 0.00 & 0.00 & 0.54 & 0.14 & 0.11 & 0.39 & 1.31 & 1.30 & 0.36 & 0.00 & 1.77 & 2.29 & 1.68 & 13.12 & 17.81 \\
        & $M_{\text{ti}} (\%)$   & 0.16 & 0.47 & 0.14 & 0.13 & 0.14 & 0.00 & 0.00 & 0.48 & 0.00 & 0.49 & 0.27 & 0.16 & 0.30 & 1.16 & 2.66 & 2.97 \\
        \midrule
        
         \multirow{6}{*}{\shortstack{ColQwen2.5\\-v0.2}} 
        & $M_{\text{cr}} (\%)$   & 59.64 & 65.89 & 56.85 & 56.64 & 58.10 & 46.41 & 57.42 & 59.93 & 62.52 & 54.85 & 56.50 & 54.11 & 31.71 & 28.13 & 16.11 & 25.03 \\
        & $M_{\text{orlb}} (\%)$ & 20.42 & 17.21 & 19.64 & 25.77 & 22.94 & 33.97 & 24.44 & 25.09 & 19.33 & 24.51 & 27.50 & 21.42 & 38.72 & 43.23 & 44.85 & 18.58 \\
        & $M_{\text{or}} (\%)$   & 18.63 & 15.50 & 21.16 & 14.77 & 14.29 & 11.81 & 14.85 & 11.77 & 13.36 & 17.23 & 12.86 & 14.01 & 9.60 & 8.39 & 5.98 & 7.61 \\
        & $M_{\text{cdr}} (\%)$  & 1.31 & 1.40 & 1.94 & 2.42 & 3.98 & 5.80 & 3.15 & 2.85 & 3.63 & 2.79 & 2.60 & 6.28 & 1.83 & 5.42 & 11.63 & 26.32 \\
        & $M_{\text{lb}} (\%)$   & 0.00 & 0.00 & 0.00 & 0.27 & 0.41 & 1.79 & 0.13 & 0.24 & 0.52 & 0.24 & 0.41 & 3.54 & 17.68 & 13.42 & 19.27 & 15.87 \\
        & $M_{\text{ti}} (\%)$   & 0.00 & 0.00 & 0.41 & 0.13 & 0.27 & 0.21 & 0.00 & 0.12 & 0.65 & 0.36 & 0.14 & 0.64 & 0.46 & 1.42 & 2.16 & 6.58 \\
        \midrule

         \multirow{6}{*}{\shortstack{GME-Qwen2\\-2B-Instruct}} 
       & $M_{\text{cr}} (\%)$   & 70.75 & 71.32 & 67.77 & 61.61 & 69.51 & 63.92 & 62.16 & 60.88 & 61.09 & 57.89 & 59.37 & 60.87 & 55.79 & 53.68 & 31.73 & 31.35 \\
        & $M_{\text{orlb}} (\%)$ & 12.25 & 12.71 & 15.35 & 20.13 & 11.95 & 19.20 & 17.87 & 22.95 & 21.92 & 22.69 & 24.21 & 19.00 & 25.61 & 18.19 & 20.60 & 13.81 \\
        & $M_{\text{or}} (\%)$   & 14.05 & 13.80 & 15.21 & 12.48 & 13.74 & 13.61 & 13.80 & 12.25 & 11.67 & 15.90 & 12.45 & 11.92 & 10.98 & 14.32 & 6.98 & 6.32 \\
        & $M_{\text{cdr}} (\%)$  & 2.94 & 1.86 & 1.52 & 3.89 & 4.26 & 2.85 & 4.73 & 2.97 & 3.63 & 2.67 & 2.05 & 5.64 & 4.27 & 10.19 & 24.42 & 28.90 \\
        & $M_{\text{lb}} (\%)$   & 0.00 & 0.00 & 0.00 & 0.94 & 0.41 & 0.21 & 1.05 & 0.48 & 1.17 & 0.49 & 1.50 & 1.29 & 2.13 & 2.45 & 11.79 & 13.29 \\
        & $M_{\text{ti}} (\%)$   & 0.00 & 0.31 & 0.14 & 0.94 & 0.14 & 0.21 & 0.39 & 0.48 & 0.52 & 0.36 & 0.41 & 1.29 & 1.22 & 1.16 & 4.49 & 6.32 \\
        \midrule

        \multirow{6}{*}{\shortstack{GME-Qwen2\\-7B-Instruct}} 
        & $M_{\text{cr}} (\%)$   & 67.81 & 68.68 & 68.88 & 61.34 & 67.58 & 62.97 & 63.34 & 58.86 & 59.66 & 54.61 & 55.54 & 60.39 & 56.71 & 60.26 & 37.87 & 34.84 \\
        & $M_{\text{orlb}} (\%)$ & 11.60 & 11.63 & 12.31 & 22.01 & 16.35 & 23.00 & 20.24 & 25.68 & 26.98 & 28.03 & 30.37 & 21.74 & 28.51 & 17.42 & 23.92 & 17.81 \\
        & $M_{\text{or}} (\%)$   & 17.81 & 16.90 & 16.04 & 11.54 & 13.19 & 10.55 & 12.75 & 11.65 & 10.89 & 14.20 & 11.63 & 10.79 & 9.76 & 12.39 & 8.47 & 5.94 \\
        & $M_{\text{cdr}} (\%)$  & 2.78 & 2.48 & 2.63 & 4.03 & 2.75 & 2.95 & 3.15 & 2.85 & 1.82 & 2.79 & 1.92 & 5.48 & 3.81 & 8.65 & 18.94 & 25.03 \\
        & $M_{\text{lb}} (\%)$   & 0.00 & 0.16 & 0.00 & 0.67 & 0.14 & 0.42 & 0.39 & 0.71 & 0.52 & 0.36 & 0.55 & 1.13 & 0.91 & 0.90 & 7.81 & 11.48 \\
        & $M_{\text{ti}} (\%)$   & 0.00 & 0.16 & 0.14 & 0.40 & 0.00 & 0.11 & 0.13 & 0.24 & 0.13 & 0.00 & 0.00 & 0.48 & 0.30 & 0.39 & 2.99 & 4.90 \\
        \midrule
        
         \multirow{6}{*}{Jina-E-v4} 
        & $M_{\text{cr}} (\%)$   & 61.76 & 64.81 & 61.13 & 60.67 & 58.65 & 57.38 & 56.64 & 54.10 & 58.88 & 57.28 & 59.92 & 60.06 & 57.16 & 55.74 & 38.54 & 36.39 \\
        & $M_{\text{orlb}} (\%)$ & 16.67 & 16.12 & 19.92 & 23.76 & 22.25 & 22.57 & 23.13 & 25.56 & 24.51 & 20.75 & 24.62 & 16.91 & 23.48 & 18.84 & 17.44 & 11.35 \\
        & $M_{\text{or}} (\%)$   & 19.77 & 17.21 & 17.15 & 13.69 & 15.66 & 18.04 & 17.08 & 17.12 & 14.79 & 20.02 & 14.36 & 18.52 & 16.01 & 17.55 & 9.97 & 8.77 \\
        & $M_{\text{cdr}} (\%)$  & 1.63 & 1.71 & 1.38 & 1.34 & 3.16 & 1.79 & 2.63 & 2.73 & 1.43 & 1.70 & 0.96 & 3.22 & 2.59 & 5.68 & 20.10 & 28.39 \\
        & $M_{\text{lb}} (\%)$   & 0.00 & 0.00 & 0.00 & 0.27 & 0.14 & 0.11 & 0.13 & 0.24 & 0.39 & 0.12 & 0.14 & 0.64 & 0.46 & 1.55 & 8.64 & 9.55 \\
        & $M_{\text{ti}} (\%)$   & 0.16 & 0.16 & 0.41 & 0.27 & 0.14 & 0.11 & 0.39 & 0.24 & 0.00 & 0.12 & 0.00 & 0.64 & 0.30 & 0.65 & 5.32 & 5.55 \\
        
        \bottomrule
    \end{tabular}
    \caption{This table presents the results for the Correct, Object Relevant Language-biased, Object Relevant, Cultural Descriptor Relevant, and Totally Irrelevant Win Percentages across various multimodal retrievers for a selection of countries.}
    \label{tab:wins_all_model_RQ3}
\end{table*}
\section{Analysis of Comparative Association Bias in Text-to-Image Retrieval}
\label{apd:sigtest}
To rigorously validate the failure modes and perspectival biases identified in RQ2 and RQ3, we employed a statistical evaluation targeting retrieval performance, embedding confusion, and the impact of query augmentation.

\subsection{Rank-1 Win Rate Analysis}
Complementing the continuous score analyses, we evaluated the model's explicit retrieval preference at the top rank using a \emph{Chi-Square ($\chi^2$) Goodness of Fit Test}. 
This approach determines whether the model exhibits a statistically significant tendency to select one specific failure mode over another when the ground truth is not prioritized. 
We defined the Null Hypothesis as the condition where the model is equally likely to retrieve either of the two competing distractor categories ($P(A) = P(B)$). 
To compute the statistic, the expected counts ($E$) for the two competing groups were derived by averaging their combined observed wins ($E_A = E_B = \frac{O_A + O_B}{2}$), while the count of unrelated or ground truth retrievals remained fixed as observed values. 
The resulting $\chi^2$ statistic, calculated with 1 degree of freedom, identifies whether the observed disparity in win rates is a result of systematic algorithmic bias.

\subsection{Test Pairs}
\label{app:test_pairs}

We conducted statistical analysis on specific pairs of candidate images, identified by the color codes in Figure \ref{fig:3query}, to isolate distinct levels of bias.
The pairs were selected to evaluate how the model resolves conflicts between semantic content, explicit cultural descriptors, and implicit language priors.

\begin{itemize}
    \item \textbf{Query Language Association Bias Test (Object-Relevant Language Bias (Purple) vs. Object Relevant (Blue))} \\
    This comparison investigates association bias in scenarios where the model successfully retrieves the correct semantic object (e.g., a train) but fails to adhere to the explicit cultural descriptor.
    By comparing a candidate that aligns with the query's language culture (Purple) against a random candidate (Blue), we test whether the script of the query exerts a stronger influence on visual selection than the explicit textual description.

    \item \textbf{Cultural Descriptor Relevance (Yellow) vs. Language Bias (Pink):} \\
    This comparison analyzes the model's prioritization mechanism during \textit{Semantic Failure}. 
    In cases where the model fails to retrieve the correct object entirely, we determine whether the retrieval is driven by a specific token match (e.g., the word "Japanese" leading to Japanese food; Yellow) or by the latent prior of the query language itself (e.g., Thai script leading to Thai cultural imagery; Pink). 
    This reveals whether the error stems from specific keyword fixation or broad language-modality leakage.
\end{itemize}

\subsection{Statistical Test Results}
Table \ref{tab:sigtest_winrate} summarizes the overall results of the statistical significance tests. For a breakdown of each test, Tables \ref{tab:sigtest_blue_purple} and \ref{tab:sigtest_yellow_pink} provide the detailed $p$-values for each model across all evaluated countries.
In the \textbf{Query Language Association Bias Test}, the majority of architectures, specifically the Vision-Language Contrastive models and LLM-Based Retrieval Embedding models, exhibited a statistically significant preference for the \emph{Object-Relevant Language Bias} candidate (Purple) over the neutral alternative (Blue).
This consistent rejection of the null hypothesis for these models confirms the presence of a strong association bias, demonstrating that the script of the input query affects the retrieved image result.
However, the Cross-lingual Alignment models (XLM-R) served as an exception, showing no statistically significant difference ($p > 0.05$) between the candidates, suggesting these models are less susceptible to implicit script bias when object relevance remains constant.
Conversely, the \textbf{Cultural Descriptor vs. Language Bias} comparison reveals a divergence in model behavior during semantic failure.
The LLM-Based Retrieval Embeddings models and Cross-lingual Alignment models displayed a statistically significant preference for the \emph{Cultural Descriptor Relevance} candidate (Yellow), indicating a general tendency to prioritize specific explicit tokens (e.g., "Japanese") over the implicit language prior.
Notably, standard CLIP-L/14 exhibited a statistically significant preference for the \emph{Language-biased category} (Pink) over the Cultural Descriptor (Diff $< 0$), suggesting that within CLIP's embedding space, the implicit bias derived from the query language script outweighs the explicit semantic cultural tokens.

\begin{table*}[htbp]
    \centering
    \small
    \setlength{\tabcolsep}{2.5pt} 
    \begin{tabular}{lcccccccccc}
    \toprule
    \multirow{3}[2]{*}{\textbf{Model}} & \multicolumn{5}{c}{\textbf{QL Association}} & \multicolumn{5}{c}{\textbf{CD vs QL}} \\
     & \multicolumn{5}{c}{\textbf{(Blue vs Purple)}} & \multicolumn{5}{c}{\textbf{(Yellow vs Pink)}} \\
    \cmidrule(lr){2-6} \cmidrule(lr){7-11}
     & \textbf{Blue} & \textbf{Purple} & \textbf{Diff} & \textbf{$p$-val} & \textbf{Sig.} & \textbf{Yellow} & \textbf{Pink} & \textbf{Diff} & \textbf{$p$-val} & \textbf{Sig.} \\
    \midrule
    \multicolumn{11}{l}{\textbf{Vision-Language Contrastive Models}}\\
    \hspace{5pt}CLIP-L/14 & 0.053 & 0.298 & -0.245 & 0.000 & True & 0.085 & 0.126 & -0.040 & 0.000 & True \\
    \hspace{5pt}CN-CLIP-L/14 & 0.098 & 0.245 & -0.147 & 0.000 & True & 0.100 & 0.112 & -0.012 & 0.006 & True \\
    \midrule
    \multicolumn{11}{l}{\textbf{Cross-lingual Alignment Models}}\\
    \hspace{5pt}XLM-R-L/14 & 0.068 & 0.072 & -0.004 & 0.290 & False & 0.076 & 0.003 & 0.073 & 0.000 & True \\
    \hspace{5pt}XLM-R-B/16plus & 0.097 & 0.099 & -0.002 & 0.647 & False & 0.066 & 0.004 & 0.062 & 0.000 & True \\
    \midrule
    \multicolumn{11}{l}{\textbf{LLM-Based Retrieval Embedders}}\\
    \hspace{5pt}ColQwen2.5-3B-M & 0.093 & 0.207 & -0.114 & 0.000 & True & 0.094 & 0.013 & 0.081 & 0.000 & True \\
    \hspace{5pt}ColQwen2.5-7B-M & 0.078 & 0.177 & -0.099 & 0.000 & True & 0.098 & 0.011 & 0.087 & 0.000 & True \\
    \hspace{5pt}ColQwen2.5-v0.2 & 0.126 & 0.201 & -0.076 & 0.000 & True & 0.096 & 0.021 & 0.074 & 0.000 & True \\
    \hspace{5pt}GME-Qwen2-2B & 0.115 & 0.184 & -0.069 & 0.000 & True & 0.081 & 0.016 & 0.065 & 0.000 & True \\
    \hspace{5pt}GME-Qwen2-7B & 0.108 & 0.208 & -0.100 & 0.000 & True & 0.068 & 0.014 & 0.054 & 0.000 & True \\
    \hspace{5pt}Jina-E-v4 & 0.136 & 0.196 & -0.060 & 0.000 & True & 0.071 & 0.012 & 0.059 & 0.000 & True \\
    \bottomrule
    \end{tabular}
    \caption{Win rates, differences, and statistical significance for the Object Relevance and Cultural Descriptor studies. Results are significant if $p < 0.05$. (Blue is the Object Relevant category, matching only the object. Purple represents the Object Relevant Language-biased category, which matches both the object and the culture of the language of the text query. Yellow represents the Cultural Descriptor Relevant category, which matches the CD's culture, and Pink is the Language-biased category, which matches only the culture of the language of the text query.)}
    \label{tab:sigtest_winrate}
    \vspace{5em}
\end{table*}
\begin{table*}[!h]
    \centering
    \footnotesize 
    \setlength{\tabcolsep}{1.5pt} 
    \begin{tabular}{@{}ll*{16}{c}@{}}
        \toprule
        \textbf{Model}& \textbf{Metrics} & \multicolumn{16}{c}{\textbf{Country}} \\
        \cmidrule(lr){3-18}
        & & \textbf{US} & \textbf{GB} & \textbf{AU} & \textbf{DE} & \textbf{CN} & \textbf{JP} & \textbf{FR} & \textbf{ES} & \textbf{AR} & \textbf{PT} & \textbf{BR} & \textbf{SA} & \textbf{TH} & \textbf{IN} & \textbf{KE} & \textbf{NG} \\
        \midrule
        
        \multirow{2}{*}{CLIP-L/14} 
        & Diff & -0.01 & 0.01 & 0.02 & -0.18 & -0.45 & -0.23 & -0.19 & -0.14 & -0.23 & -0.28 & -0.36 & -0.49 & -0.58 & -0.53 & -0.21 & -0.09 \\
        & $p$-val & \textbf{0.60} & \textbf{0.43} & \textbf{0.10} & 0.00 & 0.00 & 0.00 & 0.00 & 0.00 & 0.00 & 0.00 & 0.00 & 0.00 & 0.00 & 0.00 & 0.00 & 0.00 \\
        \midrule
        
        \multirow{2}{*}{CN-CLIP-L/14} 
        & Diff & 0.00 & -0.02 & -0.01 & -0.14 & 0.06 & -0.08 & -0.23 & -0.14 & -0.13 & -0.17 & -0.22 & -0.41 & -0.19 & -0.39 & -0.25 & -0.07 \\
        & $p$-val & \textbf{1.00} & \textbf{0.30} & \textbf{0.74} & 0.00 & 0.00 & 0.00 & 0.00 & 0.00 & 0.00 & 0.00 & 0.00 & 0.00 & 0.00 & 0.00 & 0.00 & 0.00 \\
        \midrule

        \multirow{2}{*}{Jina-E-v4} 
        & Diff & 0.00 & -0.01 & -0.08 & -0.12 & -0.01 & -0.05 & -0.09 & -0.05 & -0.10 & -0.05 & -0.10 & -0.02 & -0.07 & -0.05 & -0.06 & -0.07 \\
        & $p$-val & \textbf{0.94} & \textbf{0.73} & 0.00 & 0.00 & \textbf{0.65} & 0.01 & 0.00 & 0.03 & 0.00 & 0.02 & 0.00 & \textbf{0.39} & 0.00 & 0.01 & 0.00 & 0.00 \\
        \midrule

        \multirow{2}{*}{XLM-R-L/14} 
        & Diff & 0.02 & 0.01 & 0.03 & 0.01 & 0.00 & 0.01 & 0.00 & 0.00 & -0.04 & -0.06 & -0.08 & 0.01 & 0.00 & 0.02 & 0.02 & 0.01 \\
        & $p$-val & \textbf{0.20} & \textbf{0.33} & 0.03 & \textbf{0.32} & \textbf{0.85} & \textbf{0.70} & \textbf{0.77} & \textbf{0.72} & 0.01 & 0.00 & 0.00 & \textbf{0.38} & \textbf{0.77} & \textbf{0.21} & \textbf{0.17} & \textbf{0.39} \\
        \midrule

        \multirow{2}{*}{XLM-R-B/16plus} 
        & Diff & 0.02 & 0.04 & 0.02 & 0.02 & 0.03 & 0.02 & -0.01 & -0.04 & -0.07 & -0.08 & -0.04 & 0.04 & 0.00 & 0.03 & 0.01 & 0.01 \\
        & $p$-val & \textbf{0.24} & 0.02 & \textbf{0.26} & \textbf{0.27} & 0.06 & \textbf{0.28} & \textbf{0.42} & 0.02 & 0.00 & 0.00 & 0.03 & 0.02 & \textbf{0.80} & \textbf{0.14} & \textbf{0.48} & \textbf{0.17} \\
        \midrule
        
        \multirow{2}{*}{ColQwen2.5-3B-M} 
        & Diff & -0.03 & 0.01 & 0.00 & -0.16 & -0.03 & -0.16 & -0.22 & -0.15 & -0.15 & -0.13 & -0.20 & -0.06 & -0.18 & -0.18 & -0.10 & -0.01 \\
        & $p$-val & 0.07 & \textbf{0.69} & \textbf{0.79} & 0.00 & \textbf{0.14} & 0.00 & 0.00 & 0.00 & 0.00 & 0.00 & 0.00 & 0.00 & 0.00 & 0.00 & 0.00 & \textbf{0.51} \\
        \midrule

        \multirow{2}{*}{ColQwen2.5-7B-M} 
        & Diff & 0.02 & 0.00 & 0.03 & -0.13 & -0.06 & -0.11 & -0.23 & -0.17 & -0.15 & -0.11 & -0.16 & -0.05 & -0.17 & -0.12 & -0.09 & -0.01 \\
        & $p$-val & \textbf{0.21} & \textbf{0.92} & 0.09 & 0.00 & 0.00 & 0.00 & 0.00 & 0.00 & 0.00 & 0.00 & 0.00 & 0.01 & 0.00 & 0.00 & 0.00 & \textbf{0.45} \\
        \midrule

        \multirow{2}{*}{ColQwen2.5-v0.2} 
        & Diff & -0.05 & -0.03 & -0.02 & -0.05 & -0.02 & -0.10 & -0.13 & -0.05 & -0.06 & -0.03 & -0.13 & -0.01 & -0.16 & -0.15 & -0.18 & -0.04 \\
        & $p$-val & 0.05 & \textbf{0.13} & \textbf{0.28} & 0.01 & \textbf{0.33} & 0.00 & 0.00 & 0.01 & 0.00 & \textbf{0.12} & 0.00 & \textbf{0.76} & 0.00 & 0.00 & 0.00 & 0.01 \\
        \midrule

        \multirow{2}{*}{GME-Qwen2-2B} 
        & Diff & 0.02 & 0.01 & 0.00 & -0.07 & 0.00 & -0.09 & -0.09 & -0.11 & -0.09 & -0.07 & -0.18 & -0.10 & -0.08 & -0.06 & -0.13 & -0.03 \\
        & $p$-val & \textbf{0.41} & \textbf{0.64} & \textbf{0.89} & 0.00 & \textbf{0.94} & 0.00 & 0.00 & 0.00 & 0.00 & 0.00 & 0.00 & 0.00 & 0.00 & 0.00 & 0.00 & 0.03 \\
        \midrule

        \multirow{2}{*}{GME-Qwen2-7B} 
        & Diff & 0.05 & 0.03 & 0.02 & -0.12 & -0.01 & -0.11 & -0.14 & -0.14 & -0.11 & -0.12 & -0.24 & -0.15 & -0.14 & -0.09 & -0.16 & -0.13 \\
        & $p$-val & 0.03 & \textbf{0.13} & \textbf{0.29} & 0.00 & \textbf{0.50} & 0.00 & 0.00 & 0.00 & 0.00 & 0.00 & 0.00 & 0.00 & 0.00 & 0.00 & 0.00 & 0.00 \\
        
        \bottomrule
    \end{tabular}
    \caption{Win rate differences and $p$-values for each model across different countries in the QL Association Test.}
    \label{tab:sigtest_blue_purple}
\end{table*}
\begin{table*}[!h]
    \centering
    \footnotesize 
    \setlength{\tabcolsep}{1.5pt} 
    \begin{tabular}{@{}ll*{16}{c}@{}}
        \toprule
        \textbf{Model}& \textbf{Metrics} & \multicolumn{16}{c}{\textbf{Country}} \\
        \cmidrule(lr){3-18}
        & & \textbf{US} & \textbf{GB} & \textbf{AU} & \textbf{DE} & \textbf{CN} & \textbf{JP} & \textbf{FR} & \textbf{ES} & \textbf{AR} & \textbf{PT} & \textbf{BR} & \textbf{SA} & \textbf{TH} & \textbf{IN} & \textbf{KE} & \textbf{NG} \\
        \midrule
        
        \multirow{2}{*}{CLIP-L/14} 
        & Diff & 0.04 & 0.04 & 0.04 & 0.06 & -0.24 & 0.07 & 0.05 & 0.04 & 0.04 & 0.03 & 0.01 & -0.46 & -0.38 & -0.41 & 0.10 & 0.23 \\
        & $p$-val & 0.00 & 0.00 & 0.00 & 0.00 & 0.00 & 0.00 & 0.00 & 0.00 & 0.00 & 0.01 & \textbf{0.67} & 0.00 & 0.00 & 0.00 & 0.00 & 0.00 \\
        \midrule
        
        \multirow{2}{*}{CN-CLIP-L/14} 
        & Diff & 0.05 & 0.05 & 0.05 & 0.08 & 0.05 & 0.01 & 0.04 & 0.06 & 0.04 & 0.03 & 0.02 & -0.39 & -0.24 & -0.34 & 0.05 & 0.19 \\
        & $p$-val & 0.00 & 0.00 & 0.00 & 0.00 & 0.00 & \textbf{0.22} & 0.00 & 0.00 & 0.00 & 0.06 & \textbf{0.26} & 0.00 & 0.00 & 0.00 & 0.05 & 0.00 \\
        \midrule

        \multirow{2}{*}{Jina-E-v4} 
        & Diff & 0.02 & 0.02 & 0.02 & 0.04 & 0.05 & 0.05 & 0.04 & 0.03 & 0.02 & 0.03 & 0.02 & 0.04 & 0.02 & 0.12 & 0.18 & 0.25 \\
        & $p$-val & 0.00 & 0.00 & 0.00 & 0.00 & 0.00 & 0.00 & 0.00 & 0.00 & 0.00 & 0.00 & 0.01 & 0.00 & 0.00 & 0.00 & 0.00 & 0.00 \\
        \midrule

        \multirow{2}{*}{XLM-R-L/14} 
        & Diff & 0.05 & 0.05 & 0.06 & 0.04 & 0.04 & 0.05 & 0.04 & 0.04 & 0.05 & 0.05 & 0.04 & 0.06 & 0.06 & 0.05 & 0.09 & 0.40 \\
        & $p$-val & 0.00 & 0.00 & 0.00 & 0.00 & 0.00 & 0.00 & 0.00 & 0.00 & 0.00 & 0.00 & 0.00 & 0.00 & 0.00 & 0.00 & 0.00 & 0.00 \\
        \midrule

        \multirow{2}{*}{XLM-R-B/16plus} 
        & Diff & 0.03 & 0.04 & 0.05 & 0.04 & 0.04 & 0.05 & 0.04 & 0.03 & 0.03 & 0.03 & 0.03 & 0.08 & 0.03 & 0.05 & 0.08 & 0.35 \\
        & $p$-val & 0.00 & 0.00 & 0.00 & 0.00 & 0.00 & 0.00 & 0.00 & 0.00 & 0.00 & 0.00 & 0.00 & 0.00 & 0.00 & 0.00 & 0.00 & 0.00 \\
        \midrule
        
        \multirow{2}{*}{ColQwen2.5-3B-M} 
        & Diff & 0.08 & 0.10 & 0.10 & 0.07 & 0.03 & 0.02 & 0.06 & 0.03 & 0.04 & 0.04 & 0.04 & 0.09 & 0.02 & 0.07 & 0.20 & 0.34 \\
        & $p$-val & 0.00 & 0.00 & 0.00 & 0.00 & 0.00 & 0.00 & 0.00 & 0.00 & 0.00 & 0.00 & 0.00 & 0.00 & \textbf{0.12} & 0.00 & 0.00 & 0.00 \\
        \midrule

        \multirow{2}{*}{ColQwen2.5-7B-M} 
        & Diff & 0.04 & 0.07 & 0.07 & 0.08 & 0.04 & 0.02 & 0.07 & 0.05 & 0.05 & 0.07 & 0.07 & 0.07 & 0.02 & 0.11 & 0.23 & 0.35 \\
        & $p$-val & 0.00 & 0.00 & 0.00 & 0.00 & 0.00 & 0.00 & 0.00 & 0.00 & 0.00 & 0.00 & 0.00 & 0.00 & 0.01 & 0.00 & 0.00 & 0.00 \\
        \midrule

        \multirow{2}{*}{ColQwen2.5-v0.2} 
        & Diff & 0.05 & 0.04 & 0.05 & 0.12 & 0.06 & 0.03 & 0.08 & 0.05 & 0.06 & 0.07 & 0.05 & 0.11 & 0.02 & 0.10 & 0.07 & 0.22 \\
        & $p$-val & 0.00 & 0.00 & 0.00 & 0.00 & 0.00 & 0.00 & 0.00 & 0.00 & 0.00 & 0.00 & 0.00 & 0.00 & 0.05 & 0.00 & 0.00 & 0.00 \\
        \midrule

        \multirow{2}{*}{GME-Qwen2-2B} 
        & Diff & 0.02 & 0.04 & 0.03 & 0.06 & 0.04 & 0.03 & 0.05 & 0.03 & 0.03 & 0.06 & 0.01 & 0.07 & 0.04 & 0.11 & 0.14 & 0.28 \\
        & $p$-val & 0.00 & 0.00 & 0.00 & 0.00 & 0.00 & 0.00 & 0.00 & 0.00 & 0.00 & 0.00 & \textbf{0.11} & 0.00 & 0.00 & 0.00 & 0.00 & 0.00 \\
        \midrule

        \multirow{2}{*}{GME-Qwen2-7B} 
        & Diff & 0.03 & 0.04 & 0.04 & 0.05 & 0.03 & 0.02 & 0.05 & 0.03 & 0.04 & 0.06 & 0.02 & 0.05 & 0.02 & 0.07 & 0.13 & 0.20 \\
        & $p$-val & 0.00 & 0.00 & 0.00 & 0.00 & 0.00 & 0.00 & 0.00 & 0.00 & 0.00 & 0.00 & 0.02 & 0.00 & 0.00 & 0.00 & 0.00 & 0.00 \\
        
        \bottomrule
    \end{tabular}
    \caption{Win rate differences and $p$-values for each model across different countries in the CD vs QL test.}
    \label{tab:sigtest_yellow_pink}
\end{table*}

\section{Similarity Drift}

\subsection{Similarity Drift Significance}
\label{apd:simdrift}

Table~\ref{tab:simdrift} presents the detailed quantitative breakdown of the Similarity Drift ($\Delta \text{sim}$) analysis across all evaluated languages.
In general, adding a cultural descriptor results in a positive similarity shift across most models, validating that explicit cultural context helps narrow the retrieval scope toward the target culture.
The data reveals a critical failure mode in the baseline \emph{CLIP-L/14} model regarding non-Latin scripts. 
While CLIP performs robustly for Western languages (e.g., English, German), it exhibits negligible drift for Chinese and Thai. 
This negative value implies that for certain script-heavy or linguistically complex contexts in standard CLIP, adding a cultural descriptor introduces embedding noise rather than semantic clarity, effectively pushing the retrieved result further away from the ground truth. 
In contrast, multilingual models such as \emph{XLM-R} and \emph{GME-Qwen2} maintain consistent positive drift across all tested regions, confirming their superior cross-cultural alignment capabilities.
\begin{table*}[h]
    \centering
    \label{tab:similarity_drift}
    \footnotesize 
    \setlength{\tabcolsep}{2.5pt} 
    \begin{tabular}{@{}l*{16}{c}@{}}
        \toprule
        \textbf{Model} & \multicolumn{16}{c}{\textbf{Difference by Query Language Culture}} \\
        \cmidrule(lr){2-17}
        & \textbf{US} & \textbf{GB} & \textbf{AU} & \textbf{DE} & \textbf{CN} & \textbf{JP} & \textbf{FR} & \textbf{ES} & \textbf{AR} & \textbf{PT} & \textbf{BR} & \textbf{SA} & \textbf{TH} & \textbf{IN} & \textbf{KE} & \textbf{NG} \\
        \midrule
        CLIP-L/14          & 4.89 & 4.54 & 4.68 & 4.93 & \textbf{0.64} & 3.43 & 3.55 & 4.34 & 4.51 & 3.14 & 3.02 & \textbf{0.21} & \textbf{0.18} & \textbf{-0.16} & 2.81 & 3.74 \\
        CN-CLIP-L/14       & 5.04 & 5.09 & 5.36 & 4.40 & 5.27 & 2.80 & 3.58 & 4.44 & 4.52 & 3.52 & 3.59 & \textbf{-0.51} & \textbf{0.01} & \textbf{0.02} & 3.20 & 5.61 \\
        GME-Qwen2-2B       & 7.47 & 6.41 & 7.51 & 7.21 & 7.78 & 7.08 & 7.30 & 8.31 & 9.36 & 7.32 & 7.84 & 7.26 & 8.03 & 6.96 & 6.14 & 8.15 \\
        GME-Qwen2-7B       & 3.91 & 3.62 & 3.98 & 3.48 & 7.81 & 6.29 & 1.60 & 3.41 & 4.01 & 2.10 & 2.03 & 3.77 & 4.96 & 2.17 & 1.69 & 5.36 \\
        Jina-E-v4          & 2.82 & 2.52 & 2.57 & 1.71 & \textbf{0.73} & 2.05 & 1.67 & 2.15 & 2.66 & 1.72 & 1.28 & 2.29 & 2.69 & 1.90 & 2.76 & 1.62 \\
        XLM-R-L/14         & 5.71 & 5.28 & 5.31 & 4.13 & 3.70 & 3.06 & 4.96 & 4.92 & 5.15 & 4.93 & 4.87 & 3.86 & 3.97 & 2.82 & 4.64 & 5.65 \\
        XLM-R-B/16Plus     & 9.52 & 9.10 & 9.34 & 6.57 & 5.55 & 4.40 & 8.82 & 8.05 & 8.20 & 8.18 & 7.98 & 5.41 & 5.91 & 3.92 & 8.42 & 11.77 \\
        \bottomrule
    \end{tabular}
    \caption{Mean Similarity Drift ($\Delta Sim \times 100$) by Model and Query Language Culture. The table measures the change in cosine similarity to the \emph{Correct} when a specific cultural descriptor is added to the query. Positive values indicate successful instruction following, while near-zero or negative values (bolded) indicate that the model fails to process the cultural descriptor or treats it as noise.}
    \label{tab:simdrift}
\end{table*}

\section{Association Bias Evaluation in Verbose Text Query.}

We extended the association bias evaluation to encompass scenarios both in the absence of and incorporating explicit cultural descriptors (in both RQ2 and RQ3) in the case where we \emph{increase the verbosity of the prompt without giving more object or cultural cues}.

\subsection{Association Bias in Verbose Queries without Cultural Descriptors}
\label{apd:verbose-prompt-no-cd}
\begin{figure}[h]
  \centering
  \includegraphics[width=0.8\columnwidth]{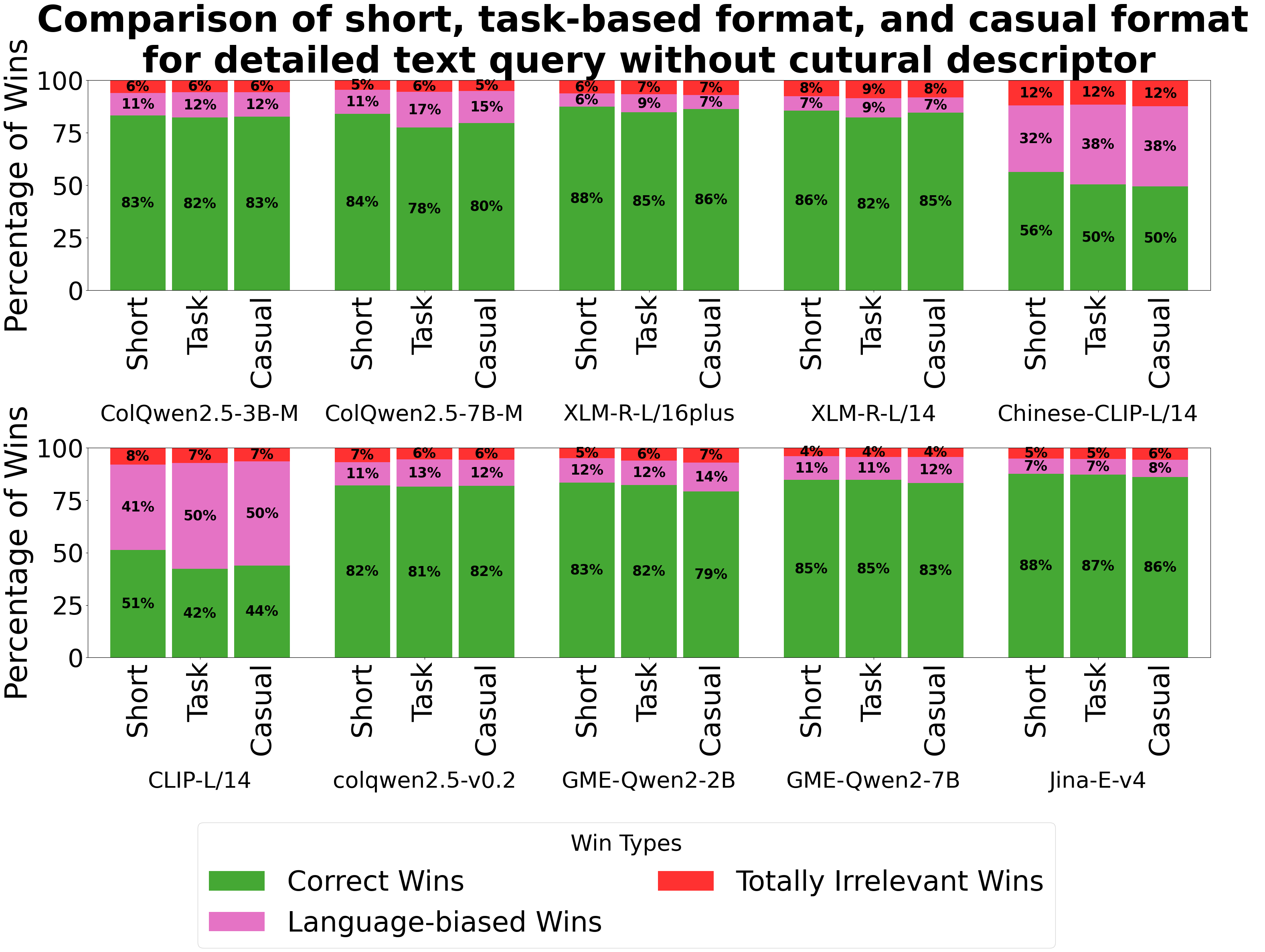}
  \caption{Comparative association bias across verbose text query in task-based and casual format without cultural descriptors.}
  \label{fig:verbose-comparison-3-candidate}
\end{figure}

\begin{figure*}[p]
  \centering
  \includegraphics[width=0.95\columnwidth]{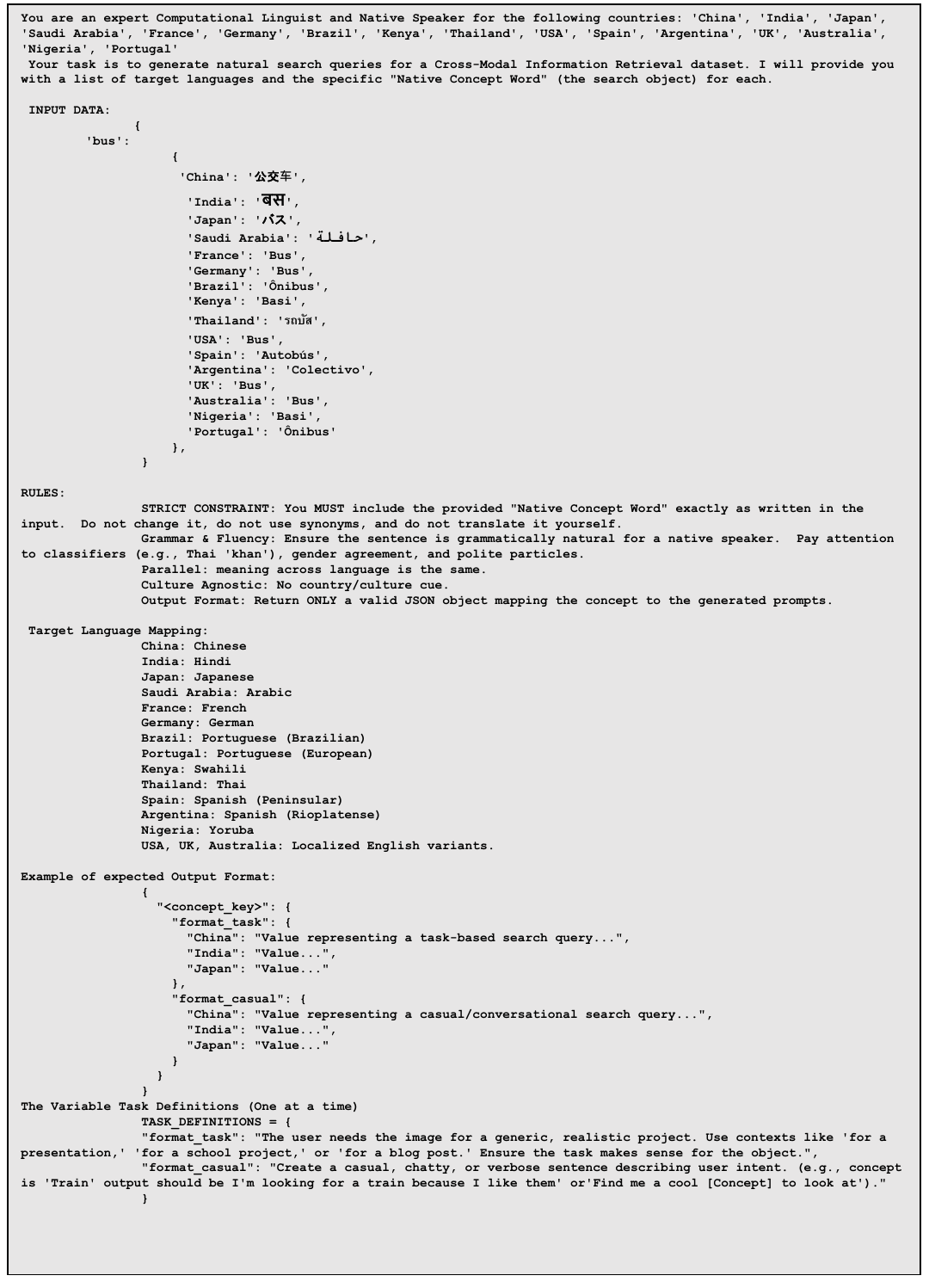}
  \caption{The prompt given to Gemini to generate unique country-specific image concepts.}
\label{fig:add_verbose_text_prompt}
\end{figure*}

Our framework evaluates association bias in detailed text queries using two primary formats: task-based and casual. The former specifies the intended use of the image (e.g., ``a picture of a train \emph{for my homework}''), whereas the latter captures the user's situational intent (e.g., ``\emph{I am looking for} a bicycle''). Query generation was facilitated by Gemini\footnote{Version used: \texttt{gemini-2.5-flash} (Released September 25, 2025). \label{fn:gemini2}}  using the prompt templates illustrated in Figure \ref{fig:add_verbose_text_prompt}. The evaluation result is shown in Figure \ref{fig:verbose-comparison-3-candidate}. The results indicate that the accuracy for verbose text queries across both task-based and casual formats decreases by 6–9\% compared to short queries (RQ2) for CLIP-L/14 and CN-CLIP-L/14. In contrast, Cross-lingual Alignment Models and MLLM-based retrieval models remain stable, with deviations of less than 4\%, except for ColQwen2.5-7B-M, which decreases by 6\% in the task-based format. The drop in accuracy occurs because the additional words in the text can distract the model from the core concept of the query. This issue is more severe in CLIP-L/14 and CN-CLIP-L/14, as these models appear to struggle with non-Latin script languages (with the exception of Chinese and Japanese for CN-CLIP-L/14). Consequently, as the query length increases, the model becomes more biased toward the culture associated with the language rather than the object.

\subsection{Association Bias in Verbose Queries with a Cultural Descriptor}
\label{apd:verbose-prompt-with-cd}
\begin{figure}[H]
  \centering
  \includegraphics[width=0.79\columnwidth]{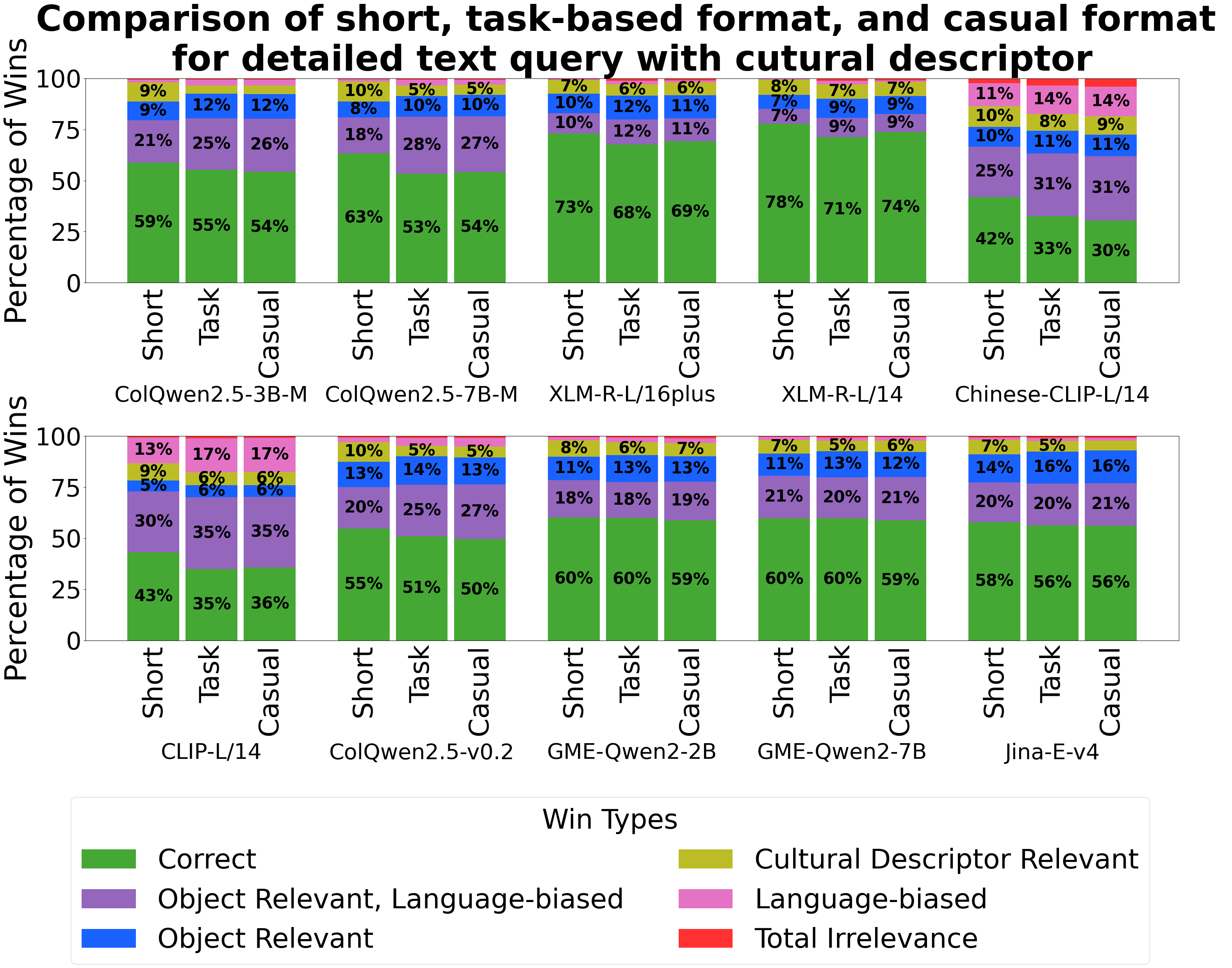}
  \caption{Comparative association bias across verbose text query in task-based and casual format with cultural descriptors.}
  \label{fig:verbose-comparison-6-candidate}
\end{figure}

To evaluate association bias in verbose queries with a cultural descriptor, a country descriptor is added to the neutral text queries from subsection \ref{apd:verbose-prompt-no-cd} using Gemini\footref{fn:gemini}, following the prompt shown in Figure \ref{fig:verbose-comparison-6-candidate}. An example of a resulting query is ``A picture of a Japanese train for my homework''. The evaluation result is shown in Figure \ref{fig:verbose-comparison-6-candidate}, short queries consistently achieve the highest correct answer across almost all models. The additional words in query cause models to drift away from the target concept or culture, especially in CLIP and CN-CLIP models, where accuracy drops by approximately 9-12\% in verbose formats. This performance loss is primarily driven by an increase in the ``Object Relevant, Language-biased'' category, indicating that verbose queries cause these models to over-rely on the language's cultural priors. Furthermore, a distinct ``Language Bias'' failure mode (Pink) is observed in CLIP models, which worsens with verbosity, confirming a tugging effect toward pure language heuristics. In contrast, the ``Cultural Diversity Relevant'' category (Yellow) remains stable (5-10\%) for most architectures but diminishes in CLIP models, suggesting that the additional text dilutes the model's focus on explicit cultural descriptors.

\begin{figure*}[p]
\centering
\begin{mdframed}[
  backgroundcolor=gray!10, 
  linewidth=1pt,           
  roundcorner=5pt,         
]
\fontsize{7.5pt}{8pt}\selectfont
\begin{verbatim}
**Role:** You are a Senior Computational Linguist and Polyglot with native-level fluency in the primary languages of the 
following countries: 
China (ZH), India (HI), Japan (JA), Saudi Arabia (AR), France (FR), Germany (DE), Brazil (PT-BR), Kenya (SW), Thailand (TH), 
USA (EN-US), Spain (ES-ES), Argentina (ES-AR), UK (EN-GB), Australia (EN-AU), Nigeria (EN-NG), and Portugal (PT-PT).

**Task:**
1. **Modify:** Take the English sentences provided in `format_task` and `format_casual`. Insert a country-specific 
reference (e.g., concept: train, mention country: China, original sentence: I need images of a Train for a school project. 
The modified sentence: I need images of a chinese train for a school project.) into the English sentence.
2. **Translate:** Translate the modified sentences into the native language of the "Target Country."

**Strict Constraints:**
1. **Grammar & Fluency:** Adjust the sentence structure, classifiers (e.g., Thai 'khan'), and gender markers to ensure 
the sentence is natural for a native speaker in that country.
2. **No Stereotypes:** Keep the sentences "Culture Agnostic." Do not add descriptors like "high-speed" for China or 
"vintage" for India. Only add the country name/adjective.
3. **Target Language Mapping:**
    *   Nigeria: Yoruba

**Input Data:**
```json
{
  "concept": "bus",
  "format_task": "I need an image of a Bus for my school project.",
  "format_casual": "I'm just looking for some cool Bus pictures to check out."
}
```

**Output Format:**
Return ONLY a valid JSON object following this exact structure:

```json
{
  "bus": {
    "format_task": {
      "Nigeria": {
        "mention_China": "Translated string here...",
        "mention_India": "Translated string here...",
        "mention_Japan": "...",
        "mention_Saudi Arabia": "...",
        "mention_France": "...",
        "mention_Germany": "...",
        "mention_Brazil": "...",
        "mention_Kenya": "...",
        "mention_Thailand": "...",
        "mention_USA": "...",
        "mention_Spain": "...",
        "mention_Argentina": "...",
        "mention_UK": "...",
        "mention_Australia": "...",
        "mention_Nigeria": "...",
        "mention_Portugal": "..."
      },

    },
    "format_casual": {
      "Nigeria": {
        "mention_China": "Translated casual string here...",
        "...": "..."
      },

    }
  }
}
\end{verbatim}
\end{mdframed}
\caption{The prompt given to Gemini to generate unique country-specific image concepts.}
\label{fig:add_cd_prompt}
\end{figure*}

\section{Semantically Focused Instruction Prompt Effects}
\label{apd:system_prompt}
\begin{figure}[!h]
  \centering
  \includegraphics[width=0.8\columnwidth]{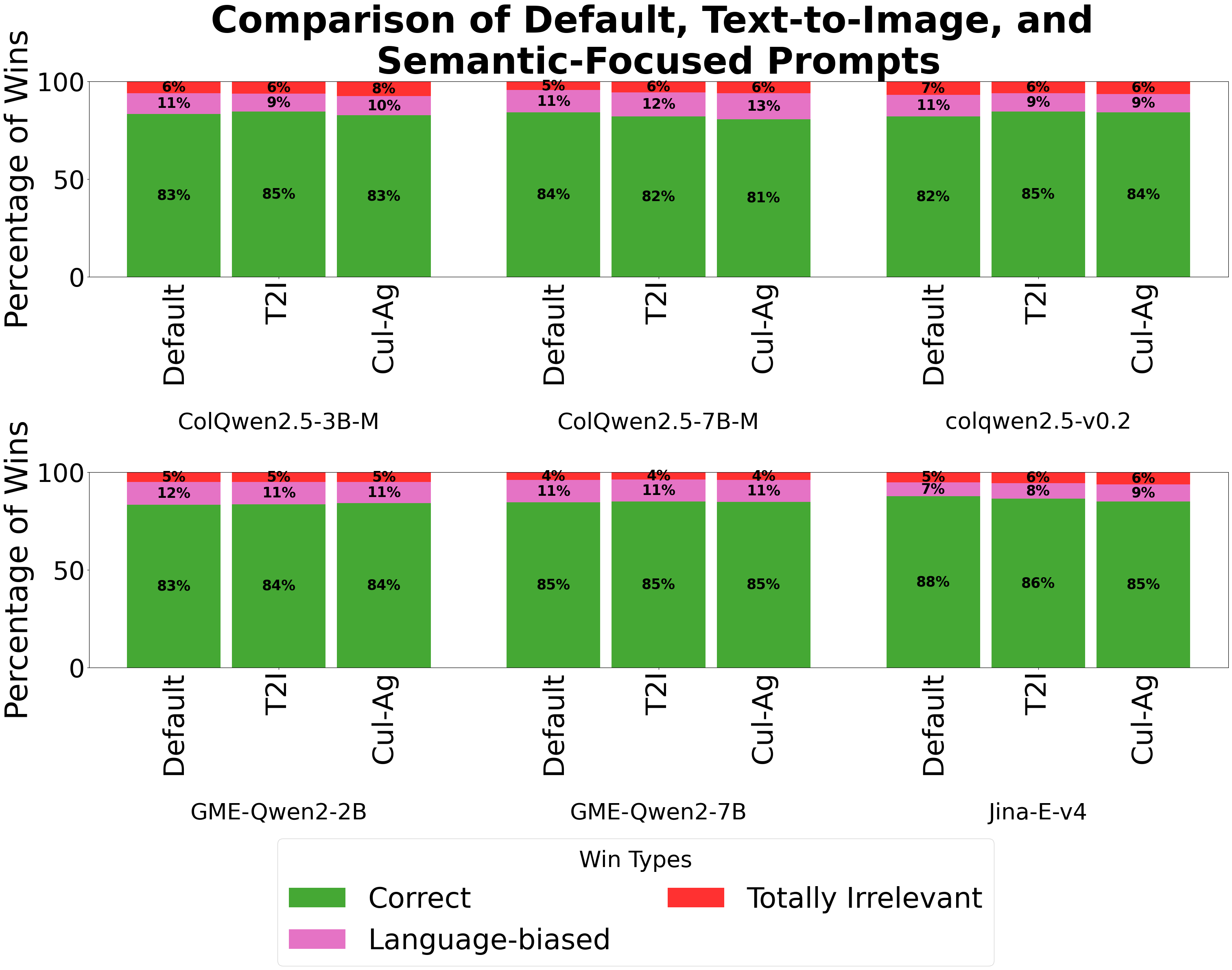}
  \caption{Comparative association bias across default prompt, standard text-to-image retriever prompt (T2I), and a culturally agnostic prompt (Cul-Ag).}
  \label{fig:compare_prompt_result}
\end{figure}

This section investigates the impact of system prompts on MLLM retriever performance. We evaluate three distinct prompts: (1) a baseline default prompt, (2) a standard text-to-image retriever prompt ("Find an image that matches the given caption"), and (3) a culturally agnostic prompt ("Find an image that matches the given caption. Focus on the semantics instead of the textual language"). The results are presented in Figure \ref{fig:compare_prompt_result}. As shown, the associative bias remains relatively consistent across different prompt configurations. This suggests that varying the system prompt does not significantly alter the model's retrieval behavior.

\section{Computational Resource}
The experiment is performed with a single A100 GPU for approximately 3 gpu hours for each model or 54 hours in total with library version of colpali-engine 0.3.13.dev1+g9bee9b2b7, transformers 4.53.3 for most experiments except GME models are utilized under transformers 4.51.3

\section{Authoring and Implementation Tools}
In preparing this manuscript, we utilized several generative large language models. For language editing and stylistic refinement, we employed Google's Gemini, along with models from xAI's Grok family (e.g., Grok Expert and Fast variants). For assistance with code implementation, scripting, and debugging, we used a model from Anthropic's Claude series (e.g., Claude Sonnet).

\end{document}